\UseRawInputEncoding
\documentclass[%
 reprint, 
 superscriptaddress,
 amsmath,amssymb,
 aps,
prb,
]{revtex4-2}
\usepackage[caption=false]{subfig} 
\usepackage{ragged2e}
\usepackage[pagewise,modulo]{lineno}
\usepackage{xcolor}
\usepackage{graphicx}
\usepackage{dcolumn}
\usepackage{bm}
\usepackage{braket}
\usepackage{subfig}
\usepackage{subcaption}
\usepackage{lipsum}
\usepackage[hidelinks,colorlinks=true,linkcolor=blue,urlcolor=blue,citecolor=blue]{hyperref}

\begin{document}

\preprint{APS/123-QED}

\title{Enhancement of Curie Temperature in \\
Ferromagnetic Insulator-Topological Insulator Heterostructures}
\author{Murod Mirzhalilov}
    \affiliation{Department of Physics, The Ohio State University, Columbus, Ohio 43210, USA}
\author{Nandini Trivedi}  
    \affiliation{Department of Physics, The Ohio State University, Columbus, Ohio 43210, USA}
\author{Mohit Randeria} 
    \affiliation{Department of Physics, The Ohio State University, Columbus, Ohio 43210, USA}

\date{\today}

\begin{abstract}
We theoretically analyze the topological insulator (TI) surface state mediated interactions between local moments in a proximate 2D ferromagnetic insulator (FMI)
motivated by recent experiments that show a significant increase in the Curie temperature $T_c$ of FMI-TI heterostructures.
Such interactions have been investigated earlier with a focus on dilute magnetic dopants in TIs.
Our problem involves a dense set of moments for which we find that the short range Bloembergen-Rowland interaction, 
arising from virtual particle-hole transitions between the valence and conduction bands, dominates over the oscillatory Ruderman-Kittel-Kasuya-Yosida (RKKY) interaction.
We show that the $T_c$ enhancement is proportional to the Van Vleck susceptibility and that the spin-momentum locking of surface states leads to 
out-of-plane ferromagnetic order in the FMI. We investigate how the hybridization between top and bottom surfaces
in a thin TI film impacts $T_c$ enhancement, and show how our results can help understand recent experiments on atomically thin Cr$_2$Te$_3$-(Bi,Sb)$_2$Te$_3$. 
\end{abstract}

\maketitle

\section{Introduction}

In recent years, the interplay between magnetism and topology has opened a new frontier in condensed matter physics, revealing novel quantum states and topological phase transitions \cite{Zhang2013, tokura2019, Bernevig_2022}. 
One of the most striking examples of this interplay occurs in magnetically doped topological insulators, where the introduction of magnetic order breaks time-reversal (TR) symmetry and opens a mass gap in the Dirac surface spectrum, leading to the realization of the quantum anomalous Hall effect (QAHE) \cite{ yu2010QAHE, ChangQAHE2013, ChangQAHE2015}.

Building on this concept, topological insulator-ferromagnetic insulator (TI-FMI) heterostructures have emerged as an interesting platform to study magnetic proximity effects without introducing disorder from doping \cite{Vobornik2011, Wei2013, Lang2014, Jiang2015}. 
In these systems, exchange coupling at the TI-FMI interface breaks TR symmetry and opens a surface gap, giving rise to QAHE-like transport signatures and magneto-optical responses \cite{TseMOKEandFaraday, Mogi_2022, Jain2024QAHE}. 
While most experimental efforts have focused on how the FMI modifies the electronic and topological properties of the TI surface states \cite{Chi2022, chang2023colloquium}, recent studies have revealed the reverse phenomenon: the influence of topological surface states on the magnetism of the adjacent FMI \cite{ou2023enhanced, CGT2014, EuS2016, YIG/TI, LCO2018, FGT2020, SRO2023}.


Our theoretical investigations are motivated by these recent observations demonstrating that TIs can exert nontrivial influences on proximate FMIs. Such effects include the induction of perpendicular magnetic anisotropy (PMA)\cite{ CGT2014, EuS2016, YIG/TI}, and, notably, the enhancement of the Curie temperature $T_c$. Representative FMI-TI heterostructures exhibiting these phenomena include Cr$_2$Ge$_2$Te$_6$-Bi$_2$Te$_3$~\cite{CGT2014}, EuS-Bi$_2$Se$_3$~\cite{EuS2016}, LaCoO$_3$-Bi$_2$Se$_3$\cite{LCO2018},
and Cr$_2$Te$_3$-(Bi,Sb)$_2$Te$_3$ \cite{ou2023enhanced}. Similar proximity-induced magnetic modifications have also been observed in ferromagnetic metal-TI systems, such as Fe$_3$GeTe$_2$-Bi$_2$Te$_3$ \cite{FGT2020}, and SrRuO$_3$-Bi$_2$Te$_3$/Sb$_2$Te$_3$ \cite{SRO2023}.

Density functional theory calculations have addressed the influence of TI surface states on the magnetic ordering in a proximate magnetic material~\cite{Kim2017DFT, FGT2020}. We take a complementary approach based on model Hamiltonians with the goal of developing a simple physical picture for $T_c$ enhancement in a 2D lattice of local moments interacting with the surface states of a TI. Since we assume the existence of local moments in the magnet, we will only look at FM insulators and we will not discuss metallic magnets. We will also focus on 2D FMIs, since for 3D magnets, it may only be the surface magnetism that is impacted. Our modeling is directly inspired by and relevant to the recent experiments on Cr$_2$Te$_3$-(Bi,Sb)$_2$Te$_3$ \cite{ou2023enhanced}, where atomically thin Cr$_2$Te$_3$ is known to be a FMI~\cite{shen2023}.

The standard theory for magnetism mediated by conduction electrons is that developed by Ruderman-Kittel-Kasuya-Yosida (RKKY) \cite{RK1954,Kasuya56,Yoshida57} which leads to the well-known oscillatory interactions with a power-law decay arising from the sharp Fermi surface. Bloembergen and Rowland (BR) ~\cite{BR1955},
showed, in a very different context, that an exchange interaction persists even in the absence of a Fermi surface, i.e. even when the chemical potential lies within a gap.
The BR mechanism, on the other hand, results in an exchange interaction between the moments that decays as a power-law at short distances and cuts off exponentially on a length scale determined by the gap. It is mediated by virtual electron-hole transitions across the gap from the filled valence band to the empty conduction band. 

These ideas have been extended to magnetically doped TIs~\cite{Zhang2009, Ye_2010, Biswas2010BR, Garate2010RKKY, Chang2011, Galitski2014, Zyuzin2014} but, as explained in greater detail below, the focus previously in the literature was on a quite different regime than the one we analyze for our problem of local moments in a FMI. Briefly, the differences are as follows. Consider two local moments in the FMI separated by ${\bf R}$. For computing the $T_c$ enhancement, we need to sum the exchange couplings over all ${\bf R}$ and obtain the ${\bf Q} = 0$ response of the susceptibility. Thus we are not just interested in the long-distance asymptotics of the interaction, as the response on the scale of the lattice spacing $a$ also plays an important role in our analysis. Furthermore, for finite doping on the TI surface states with a chemical potential that lies within the band, it is necessary to include the exchange interaction contributions between moments with separation $R \ll 1/k_F$. This is the opposite of the dilute magnetic dopants with $k_F R \gg 1$ of interest in previous works~\cite{Garate2010RKKY, Chang2011, Zyuzin2014}.

\begin{figure}[t]
  \centering
  \includegraphics[width=\linewidth]{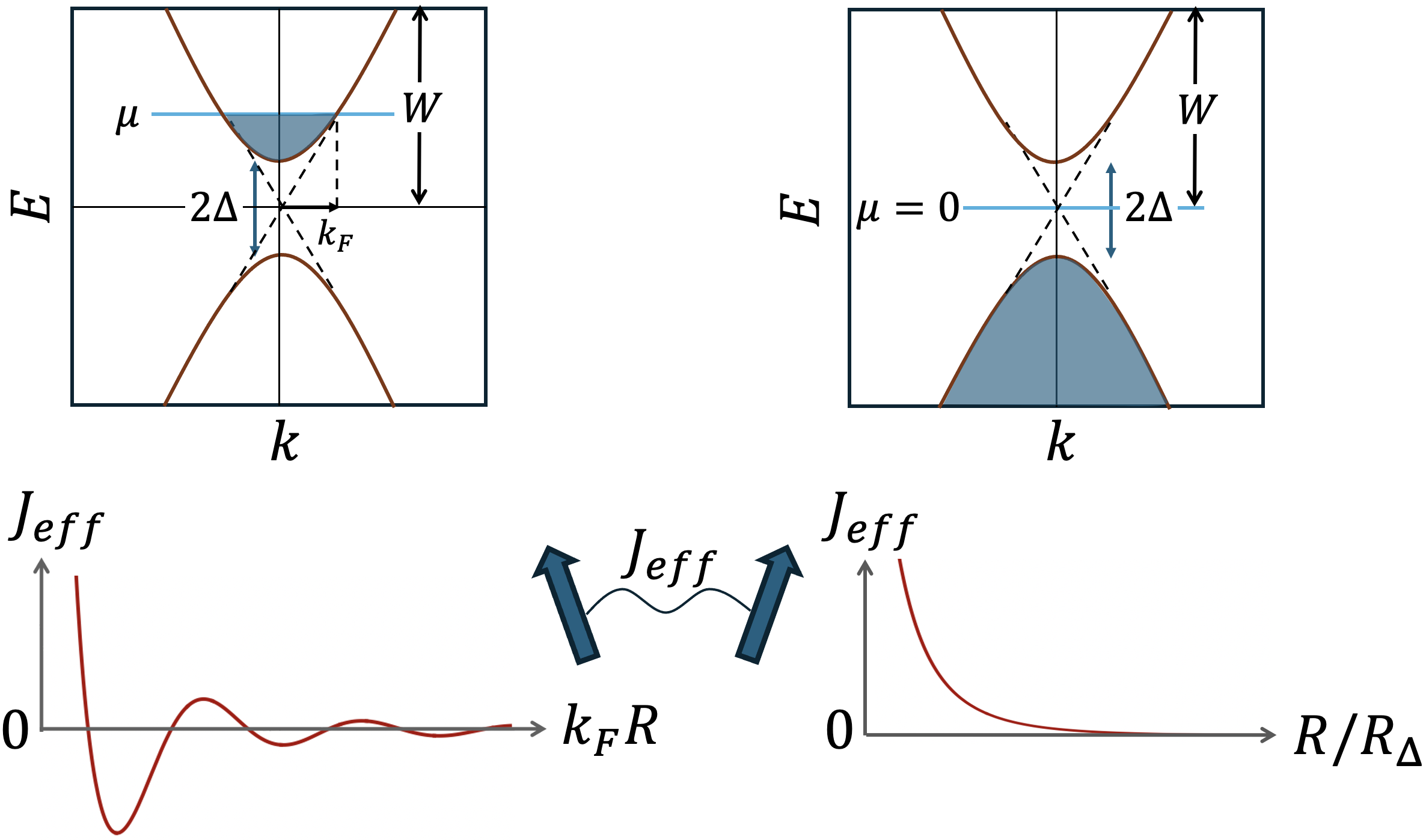}\hfill
  \caption{\justifying Schematic illustration of the effective exchange interaction between local moments mediated by Dirac electrons. (Left) When the chemical potential $\mu$ lies in the conduction band, the resulting interaction is the oscillatory RKKY type, decaying as a power law at large distances $k_F R \gg 1$. (Right) When the chemical potential lies inside the gap ($\mu=0$), the exchange is short-ranged, corresponding to the Bloembergen--Rowland (BR) interaction that decays exponentially with the characteristic length scale $R_\Delta = \hbar v_F / \Delta$. 
Here, $W$ denotes the bandwidth, and $\Delta$ is the exchange gap.}

  \label{fig:SystemDepiction}
\end{figure}

We conclude this Section by summarizing our main results and the organization of our paper. 
In Section~\ref{Sec.II} we introduce our model which describes the Dirac surface states of TI with a possible mass term arising from broken time reversal coupled to classical local moments with an exchange coupling $J$. We derive the form of the exchange interaction between two moments separated by a distance $R$.
This surface-state-mediated exchange interaction exhibits characteristic anisotropies that originate from the spin-momentum locking in the TI.

We discuss in Section~\ref{Sec.III} two contributions to the exchange interaction, one from the oscillatory RKKY interaction arising from the Fermi surface and the other from the BR interaction from virtual transitions out of the filled valence band. We show that the BR interaction dominates in the regime of interest and therefore set the chemical potential to zero, so that there is no Fermi surface.

We first show in Section~\ref{Sec.IV} how a lattice of local moments interacting with the BR interaction leads to a FM ground state with out-of-plane ordering because of
the anisotropies in the couplings. We next estimate the enhancement of $T_c$ within mean-field theory and show that it is proportional to the Van Vleck susceptibility of the surface-state electrons. 

In Section~\ref{Sec.V}, we discuss thin TI films where we need to take into account the hybridization between TI surface states. This opens up a gap in general 
but is also known to lead to topological phase transitions. We analyze how the BR coupling is impacted by these effects and predict the $T_c$ enhancement
in 3 and 4 QL (quintuple layer) (Bi,Sb)$_2$Te$_3$ films and compare them with the thick TI film results.

Finally, in Sec.\ref{Sec.VI} we compare our results with experiments on Cr$_2$Te$_3$-(Bi,Sb)$_2$Te$_3$~\cite{ou2023enhanced} and find satisfactory agreement.

\section{Model Hamiltonian and Exchange interaction}
\label{Sec.II}

We begin with the 2D problem of the surface states of a TI film interacting with the local moments of a proximate FMI. 
The low-energy Hamiltonian describing the TI surface states is given by~\cite{Hasan_20103DTI, Qi3DTI}
\begin{equation}\label{Eq:H0}
     H_0=\hbar v_F (\boldsymbol{k}\times\boldsymbol{\sigma})\cdot\hat{z}+\Delta \sigma_z,
\end{equation}
where $v_F$ is the Fermi velocity, $\boldsymbol{k}=(k_x,k_y)$ is the wavevector, and 
$\boldsymbol{\sigma}=(\sigma_x,\sigma_y,\sigma_z)$ are the Pauli spin matrices. $\Delta$ opens a gap at the Dirac point as 
a result of the broken TR symmetry of the FM order at the interface. $H_0$ describes the system on energy scales smaller than 
the bandwidth $W=\hbar v_F/a$, where $a$ is the lattice spacing.

To understand the exchange interaction mediated by the surface states, we first consider the simple problem of two (classical) magnetic moments 
$\boldsymbol{S}_i$ located at $\boldsymbol{R}_i$ ($i=1,2$) that interact with the electrons described by $H_0(\boldsymbol{k})$; 
later we will generalize to a lattice of local moments. This system is described by the Hamiltonian
\begin{equation}\label{Eq:H}
H=H_0+H_{\rm int}, \ \ \ 
H_{\rm int}=Ja^2 \sum_{i}\boldsymbol{\sigma}\cdot\boldsymbol{S}_i\, \delta(\boldsymbol{r}-\boldsymbol{R}_i),
\end{equation}
where the energy scale $J$  is an isotropic Kondo or s-d coupling between the Dirac electron spin and the local moments,
which are treated as classical spins. 

Treating $H_{\rm int}$ as a perturbation to second order in $J/W \ll 1$ we obtain
the exchange interaction between moments mediated by itinerant electrons
\begin{equation}\label{Eq:Hex}
\begin{aligned}
        \mathcal{H}_{\rm ex}=-{J^2\over W}\sum_{\alpha\beta}\chi_{\alpha\beta}(\boldsymbol{R}_1-\boldsymbol{R}_2)S_1^\alpha S_2^\beta;
\end{aligned}
\end{equation}
see Appendix \ref{App:RKKY} for details. Here $\chi_{\alpha\beta}(\boldsymbol{R})$ is the spatially-dependent static spin susceptibility
which, at $T=0$, is given by
\begin{equation}\label{Eq:chi}
    \chi_{\alpha\beta}(\boldsymbol{R})=\frac{W}{\pi}{\rm Im}\int^\mu_{-\infty}d\omega \text{Tr}[\sigma_\alpha G(\boldsymbol{R},\omega^+)\sigma_\beta G(-\boldsymbol{R},\omega^+)]
\end{equation}
where $(\alpha,\beta)=(x,y,z)$, the trace is taken over spin, $\mu$ is the chemical potential, and 
$G(\boldsymbol{R},\omega^+)$ is the retarded ($\omega^+=\omega+i0^+$) Green's function for the electrons in the TI surface states. The bandwidth factor $W$ is introduced in the above equations so that $\chi$ is dimensionless. We focus on $T=0$ here because temperature has a minimal effect on the exchange couplings 
so long as $k_BT\ll W$ as shown in Appendix \ref{APP:VVFiniteT}.

The form of $G(\boldsymbol{R},\omega^+)$ is well known \cite{Chang2011, Pesin2011, Galitski2014}, and for completeness
we discuss in Appendix \ref{App:RealSpaceGF}. Using this Eq.\eqref{Eq:Hex} can be rewritten in the form
\begin{equation}\label{Eq:FinalHex}
    \begin{aligned}
        \mathcal{H}_{ex} =&- J_{zz}(\boldsymbol{R})S^z_1S^z_2-J_\perp (\boldsymbol{R})S_1^\perp S_2^\perp-J_{\Vert}(\boldsymbol{R})S^\Vert_1S^\Vert_2\\
        &-J_{DM}(\boldsymbol{R})(\hat{\boldsymbol{R}}\times \hat{z})\cdot(\boldsymbol{S}_1\times \boldsymbol{S}_2)
    \end{aligned}
\end{equation}
where $S_i^\perp=\boldsymbol{S}_i\cdot(\hat{\boldsymbol{R}}\times \hat{z})$, $S^{\Vert}_i=\boldsymbol{S}_i\cdot\hat{\boldsymbol{R}}$ and $\hat{\boldsymbol{R}}=\boldsymbol{R}/R$, where $\boldsymbol{R} = \boldsymbol{R}_1 - \boldsymbol{R}_2$.
The first term is an Ising interaction, the next two terms involving $S_i^\perp$ and $S_i^{\Vert}$ describe anisotropic exchange couplings, 
and the last term is the Dzyaloshinskii-Moriya (DM) interaction. 
The anisotropic nature of the interactions between local moments originates from the spin-momentum locking of the TI surface state electrons \cite{Pesin2011}.

The couplings in Eq.\eqref{Eq:FinalHex} are given by
\begin{equation}\label{Eq:Exchange}
    \begin{aligned}
        J_{zz}(\boldsymbol{R}) &=\frac{J^2}{8\pi W^4}{\rm Im}\int_{-\infty}^\mu d\omega\Big[(\mathcal{E}^2-2\Delta^2)\text{H}_0^{(1)}(i\rho)^2\\
        & \ \ \ \ \ \ \ \ \ \ \ \ \ \ \ \ \ \ \ \ \ \ \ \ \ \ \ -\mathcal{E}^2\text{H}_1^{(1)}(i\rho)^2\Big]\\
       J_{\perp}(\boldsymbol{R}) &=\frac{J^2}{8\pi W^4}{\rm Im}\int_{-\infty}^\mu d\omega\mathcal{E}^2\Big[\text{H}_0^{(1)}(i\rho)^2+\text{H}_1^{(1)}(i\rho)^2\Big]\\
       J_{\Vert}(\boldsymbol{R}) &=\frac{J^2}{8\pi W^4}{\rm Im}\int_{-\infty}^\mu d\omega\mathcal{E}^2\Big[\text{H}_0^{(1)}(i\rho)^2-\text{H}_1^{(1)}(i\rho)^2\Big]\\
       J_{DM}(\boldsymbol{R}) &=\frac{J^2}{4\pi W^4}{\rm Im}\int_{-\infty}^\mu d\omega\Big[i\omega^+\mathcal{E}\text{H}_0^{(1)}(i\rho)\text{H}_1^{(1)}(i\rho)\Big]\\
    \end{aligned}
\end{equation}
Here $\text{H}^{(1)}_{0,1}$ denotes the zeroth and first order Hankel function of the first kind and
$\rho=\mathcal{E}R/\hbar v_F$ where $\mathcal{E}=\sqrt{\Delta^2-(\omega^+)^2}$.
Eqs.~\eqref{Eq:FinalHex} and \eqref{Eq:Exchange} generalize the results of ref.~\cite{Chang2011} to finite $\Delta$. 

\section{Bloembergen-Rowland versus oscillatory RKKY interactions}
\label{Sec.III}

To understand which couplings are important 
we need to first look at various length scales in our problem: the separation between magnetic moments $R$, 
the lattice constant $a$, the length scale associated with the exchange gap $R_\Delta=\hbar v_F/\Delta$, and that associated 
with the Fermi surface $1/k_F$. Note that when the chemical potential $\mu$ is inside one of the two bands 
$k_F$ is finite, but it vanishes when $\mu$ lies within the TR symmetry breaking gap $\Delta$ in the surface state dispersion.

We will see below that for the calculation of $ T_c$ enhancement, the sum over all ${\bf R}$ of the
exchange couplings, i.e., the ${\bf Q} = 0$ contribution is required.
Thus, our focus is quite different from the asymptotic large $R$ dependence that much of the literature focuses on, even though we will
discuss that as well. In particular, it is important for us to understand the {\it short distance regime} where $R \sim a \ll (R_\Delta, 1/k_F)$, with $k_F R_\Delta$ arbitrary.
We focus on $R\sim a$ since the magnetic moments in the FMI lie on a lattice with spacing $a$.
Further, the relevant gap scale $\Delta$ is such that $a\ll R_\Delta$.
For example, using $\Delta \simeq 50$ meV~\cite{Luo2013Gap} and $v_F \simeq 3.69 \times 10^5$  m/s for (Bi, Sb)$_2$Te$_3$ \cite{BST2015}, we find that 
$R_\Delta\simeq 5$nm $\gg a\simeq 0.4$ nm. This continues to be true at higher temperatures where 
$\Delta$ decreases. In fact, in our analysis of $T_c$  enhancement, we set $\Delta$ to zero (just above $T_c$), which corresponds to
a divergent $R_\Delta$.
Finally, we are in the ``low-doping regime" $a\ll 1/k_F$. In Bi$_2$Se$_3$, if $k_Fa$ is of order one, then the chemical potential
$\mu \simeq 0.6$ eV, which would be well within the {\it bulk} conduction band~\cite{Chi2022}, a situation that is not relevant for our regime of interest in which
$\mu$ either lies within the gap $\Delta$ or at most inside the surface state bands.

Analytical results have been obtained earlier for massless ($\Delta=0$) Dirac electrons for
$\mu=0$~\cite{Biswas2010BR, Garate2010RKKY, Pesin2011} and $\mu\neq 0$~\cite{Garate2010RKKY, Chang2011}.
In the latter case, the focus was on dilute magnetic ions on the surface of TIs corresponding to $k_F R  \gg 1$, exactly the opposite of
$k_F R \ll 1$ regime of our interest. Furthermore, we also need $\Delta\neq 0$ results for our $T=0$ analysis, where we cannot use 
known analytical results and must evaluate the integrals in Eq.~\eqref{Eq:Exchange} numerically.

\begin{figure}[t]
  \centering
  \includegraphics[width=0.48\linewidth]{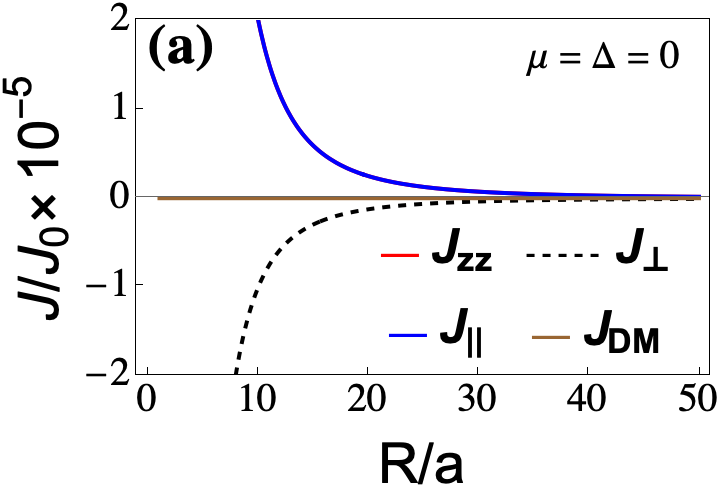}\hfill
  \includegraphics[width=0.48\linewidth]{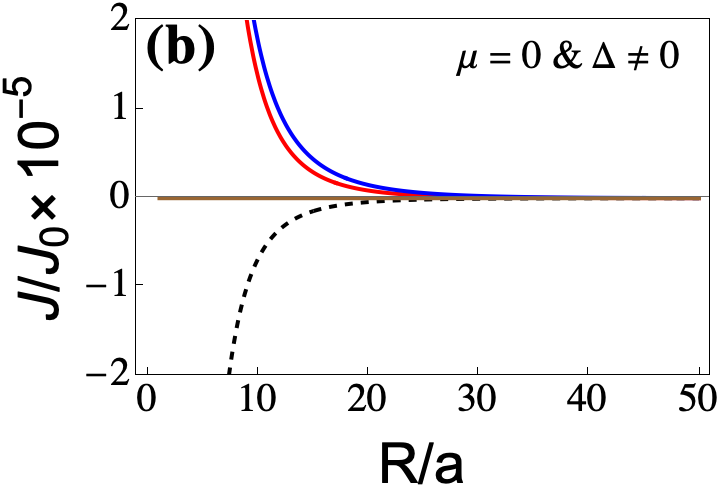}\\[4pt]
  \includegraphics[width=0.48\linewidth]{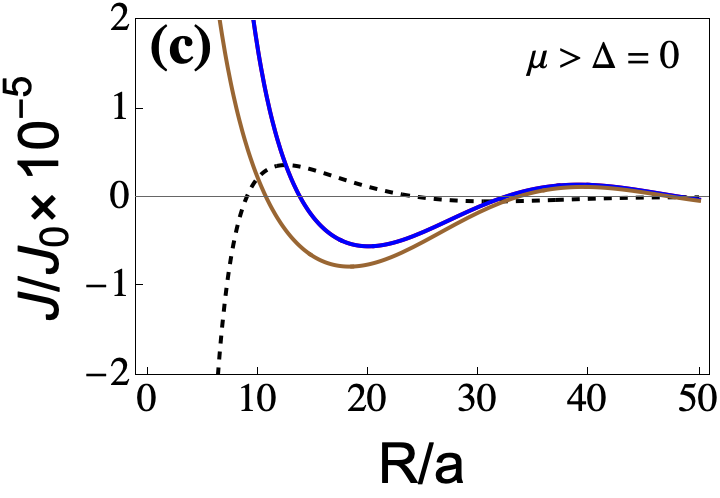}\hfill
  \includegraphics[width=0.48\linewidth]{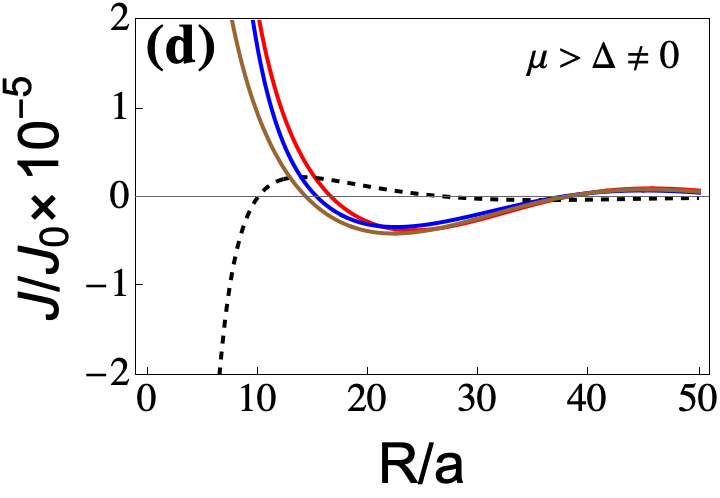}
  \caption{\justifying Coupling strengths in units of $J_0=J^2/W$ vs $R/a$. Here, $\Delta=30$ meV in panels (b) \& (d) and $\mu=60$ meV in panels (c) \& (d). We note that in panels (a) \& (c) the red and blue curves coincide since $J_{zz}=J_{\parallel}$ for $\Delta=0$.}
  \label{fig:Couplings}
\end{figure}

Before describing our calculations of the exchange interactions based on analysing Eqs.~\eqref{Eq:FinalHex} and \eqref{Eq:Exchange},
we summarize the main physical picture that emerges.
The integrals in Eq.~\eqref{Eq:Exchange} can be split into the sum of two terms: 
$\int_{-\infty}^0$ and $\int_{0}^\mu$. (We assume $\mu > 0$ here; the  $\mu < 0$ case be
handled in an analogous way.) We call the valence electrons contribution, arising from $\int_{-\infty}^0$ in Eq.~\eqref{Eq:Exchange}, 
the Bloembergen-Rowland (BR) interaction \cite{BR1955}, while that from the Fermi sea, coming from $\int_0^{\mu}$, is the usual oscillatory RKKY interaction. 
The nomenclature ``BR interaction" was historically used only for insulators but we use it in the more general sense defined here. 

We note that the BR interaction in insulators originates from virtual interband transitions. 
This is perhaps less transparent in the form in which our expressions are written, and more so if 
in evaluating the susceptibility in Eq.\eqref{Eq:Hex} we had first done a Matsubara sum in the finite temperature formalism; see Appendix \ref{App:RKKY}. However, as discussed there, this also leads to more complicated convergence issues that are avoided
in the approach that we take here. 

We show next that, in the low-doping regime $R \sim a\ll 1/k_F$, BR contribution
arising from the filled valence band dominates over the RKKY contribution from 
the Fermi surface of the partially filled conduction band. 
Let us first demonstrate this for the gapless case ($\Delta=0$) where one can analytically evaluate~\cite{Chang2011} the 
integrals in Eq.~\eqref{Eq:Exchange} in terms of Meijer-G functions; see Appendix \ref{App:nonzeromu}. 
In the regime of our interest we find that
\begin{equation}\label{Eq:smallkFR}
    \begin{aligned}
        J_{zz} &=J_\Vert \approx \frac{J_0}{16\pi (R/a)^3}\Big(1+\frac{8}{3\pi}[\gamma_1+\ln (k_FR)](k_FR)^3\Big)\\
        J_{\perp} &\approx -\frac{J_0}{32\pi (R/a)^3}\Big(1-\frac{8}{3\pi}[\gamma_2+2\ln  (k_FR)]( k_FR)^3\Big)\\
        J_{DM} &\approx \frac{J_0}{4\pi^2 (R/a)}(k_Fa)^2 \qquad \qquad\qquad (k_FR\ll 1)
    \end{aligned}
\end{equation}
where we have defined $J_0 = J^2/ W$.
The leading order contributions, obtained by setting $k_F = 0$, coincide with the results of
refs. \cite{Biswas2010BR, Garate2010RKKY, Pesin2011}, while
the first order corrections involve constants $\gamma_1\approx 0.05$ and $\gamma_2\approx -1.90$ that
arise from expanding Meijer-G functions near the origin. 
We thus see that the leading contributions to $J_{zz}, J_{\Vert}$, and $J_{\perp}$ in Eq.~\eqref{Eq:smallkFR} come from 
the valence electrons which contribute even in the $k_F =0$ limit, while the finite $k_F$ contribution arising from conduction electrons
is a small correction for $k_F R \ll 1$.
The DM interaction is much smaller than all the other exchange interactions with $J_{DM}/J_{zz} \sim (k_F R)^2$.

Next, let us turn to the case of a gapped spectrum with $\mu>\Delta$, which requires numerical evaluation of the 
integrals in Eq.~\eqref{Eq:Exchange}. Here too we find that, for $R \sim a \ll (R_\Delta, 1/k_F)$,
the BR contribution dominates over the oscillatory RKKY. We find that at $R = a$ the ratio of the BR to RKKY contributions $J_{zz}^{BR}/J_{zz}^{RKKY} \sim 10^2$ 
at $k_F a = 0.1$. It becomes even larger with decreasing $k_F a$ since $J_{zz}^{BR}$ is independent of $k_F$ while
$J_{zz}^{RKKY}$ vanishes as $k_F \to 0$.

It is useful look at the exchange interactions plotted as a function of $R/a$ in Fig.~\ref{fig:Couplings}, which 
will further clarify how the BR interaction dominates over the oscillatory RKKY interaction.
From panels (a) ($\mu=\Delta= 0$) and (b) ($\mu=0, \Delta\neq 0$), we see that the couplings are nonzero even in the absence of a Fermi surface, 
and thus the only contribution is from the BR mechanism. The results of panel (a) are equivalent to Eq.~\eqref{Eq:smallkFR} with $k_F = 0$.
With a non-zero $\Delta$ the interactions in panel (b) retain their $R^{-3}$ form for short distances $R\ll R_\Delta$, but decay exponentially 
as $\sim \exp(-2R/R_\Delta)$ at long distances $R\gg R_\Delta$. 
The presence/absence of a gap does not alter the sign of the couplings: $J_{zz}$ and $J_\Vert$ are ferromagnetic whereas 
 $J_{\perp}$ is antiferromagnetic. The DM interaction vanishes since $k_F = 0$.
 
We examine the role of a finite $k_F =\sqrt{\mu^2-\Delta^2}/\hbar v_F$ in the gapless and gapped cases in panels (c) and (d) of Fig.~\ref{fig:Couplings}
respectively. The long-distance ($k_FR\gg 1$) behavior of the exchange interactions is dominated by the Fermi surface contribution
and has the well known 2D RKKY oscillatory form $R^{-2}\cos{(2k_FR+\phi)}$; see Appendix \ref{App:nonzeromu}. 

As we shall see in the next Section, the quantity that enters our estimate for $T_c$ enhancement due to exchange mediated by TI surface states is
$J_{zz} ({\bf Q} = 0) = \sum_{\bf R} J_{zz} ({\bf R})$ which we can break up into the sum of two terms $J_{zz}^{BR} ({\bf Q} = 0) + J_{zz}^{RKKY} ({\bf Q} = 0)$,
where the BR term comes from the contribution of the filled valence band while the RKKY term comes from the conduction electrons.
Both ${\bf R}$ sums are convergent and a numerical evaluation shows that the BR term is an order of magnitude larger than the RKKY term for 
$k_F a = 0.1$ and their ratio only grows with decreasing $k_F a$. Therefore, we set $\mu = 0$ from hereon and only focus on the BR interaction. 
Recall that, in Section II we saw that the exchange interaction $J_{zz} ({\bf R}) = J_0 \chi_{zz} ({\bf R})$. The uniform susceptibility
corresponding the the BR interaction is called the {\it Van Vleck susceptibility}
$\chi_{VV} \equiv \chi_{zz}^{BR} ({\bf Q} = 0)$ \cite{VanVleck1932}.

\section{Lattice of magnetic moments}\label{Sec.IV}

Our goal is to see how the TI surface-state-mediated interactions impact the magnetism in 2D FMI film coupled to the TI.
We first look at the TI surface-state mediated interactions between local moments 
without taking into account the exchange couplings intrinsic to the FMI.  At the next stage, we
include the interactions that lead to ferromagnetism in the FMI even in the absence of the TI.

We write down the simplest 2D lattice Hamiltonian that generalizes Eq.\eqref{Eq:FinalHex}
\begin{equation}\label{Eq:BRHamiltonian}
\begin{aligned}
    \mathcal{H}_{ex} =- \frac{1}{2}\sum_{i,j} \Big[ &J_{zz}(\boldsymbol{R}_{ij})S^z_iS^z_j+J_\perp (\boldsymbol{R}_{ij})S_i^\perp S_j^\perp\\
    &+J_{\Vert}(\boldsymbol{R}_{ij})S^\Vert_iS^\Vert_j\Big]
\end{aligned} 
\end{equation}
where the classical spin-$S$ moments ${\bf S_i}$ are located at lattice sites $\boldsymbol{R}_{i}$ with $\boldsymbol{R}_{ij}=\boldsymbol{R}_{i}-\boldsymbol{R}_{j}$.  
The factor of $(1/2)$ compensates for the double counting in the sum over all lattice sites in the FMI.
We take into account only the BR interaction originating from the valence band, which we have shown 
dominates over Fermi surface effects in the regime of our interest.
These interactions are ``short-ranged" as described above with a $1/R^3$ behavior at short distances with an exponential fall-off on the 
scale of $R_\Delta = \hbar v_F/\Delta$. 

The BR exchange interactions are obtained by setting $\mu=0$ in Eq.~\eqref{Eq:Exchange}, and thus
the DM term vanishes.
We can rewrite these results in a more convenient form in terms of the zeroth- and first-order modified Bessel functions $\text{K}_{0,1}^2$ of the second kind. We find
\begin{equation}\label{Eq:BRCouplings}
    \begin{aligned}
        J_{zz}(\boldsymbol{R}) &=\frac{J_0}{2\pi^3 (R/a)^3}\int_{0}^{\infty}  dy\Big[(y^2-\xi^2)\text{K}_0^2(\sqrt{\xi^2+y^2})\\
        &+(y^2+\xi^2)\text{K}_1^2(\sqrt{\xi^2+y^2})\Big]\\
       J_{\perp}(\boldsymbol{R}) &=\frac{J_0}{2\pi^3 (R/a)^3}\int_{0}^{\infty}  dy(y^2+\xi^2)\Big[\text{K}_0^2(\sqrt{\xi^2+y^2})\\
       &-\text{K}_1^2(\sqrt{\xi^2+y^2})\Big]\\
       J_{\Vert}(\boldsymbol{R}) &=\frac{J_0}{2\pi^3 (R/a)^3}\int_{0}^{\infty}  dy(y^2+\xi^2)\Big[\text{K}_0^2(\sqrt{\xi^2+y^2})\\
       &+\text{K}_1^2(\sqrt{\xi^2+y^2})\Big]\\
    \end{aligned}
\end{equation}
where we have defined $\xi= R/R_\Delta$.
Eqs.~\eqref{Eq:BRHamiltonian} and \eqref{Eq:BRCouplings} define the problem of BR-coupled lattice of localized magnetic moments.

\subsection{Ground State and Excitation Spectrum}

Given that the Hamiltonian~\eqref{Eq:BRHamiltonian} has anisotropic interactions, it is important to
first understand the nature of its ground state. We find that the system exhibits ferromagnetic ordering at $T=0$
with the spins aligned in the $z$-direction. We briefly summarize these results here with a details relegated to
Appendix \ref{App:BRGroundState}.

First, we have exactly minimized the Hamiltonian for a small cluster of spins to find that the FM ground state.
Second, we have checked the local stability of the FM ground state by performing a spin-wave analysis.
We find that the spin wave spectrum is gapped at all $\boldsymbol{q}$, and is
of the form $\omega(\boldsymbol{q})\approx K+Aq^2$ near $\boldsymbol{q}=0$.
Here $A>0$ is the spin stiffness which we find to be essentially independent of $\Delta$,
We also find that Ising anisotropy $K=f(\Delta/W) S J_0$, where $S$ is the spin value, $J_0 = J^2/W$ and 
$f$ is a dimensionless function of order unity which shows a small quadratic reduction for small $0 \leq \Delta/W \leq 0.1$.
The easy-axis Ising anisotropy arises from the anisotropic nature of the BR interaction, which in turn has its origin in the spin-momentum
locking of the TI surface states.

\subsection{$T_c$ Enhancement}

 We now turn to the analysis of how the BR coupling enhances the $T_c$ of the FMI. We make no attempt to make a microscopic model of 
 the intrinsic ferromagnetic coupling in the FMI. Since a 2D FMI film orders, we simply model it as a nearest-neighbor Ising model 
 ${\cal H}_I = -1/2\sum_{i,j}J_I S^z_iS^z_j$ with a FM coupling $J_I$; the subscript $I$ denotes the intrinsic Ising coupling in the insulator.
 This simplification plays no role in our analysis of $T_c$ enhancement, as it only determines the intrinsic $T_c$ of the FMI.
 The full Hamiltonian is ${\cal H} = {\cal H}_I + \mathcal{H}_{ex}$, where $\mathcal{H}_{ex}$ is the TI surface state mediated BR interaction
 of Eqs.~\eqref{Eq:BRHamiltonian} and \eqref{Eq:BRCouplings}.
 
We analyze the system within a simple mean-field theory, which we recognize will give an overestimate of $T_c$, especially in 2D.
With spins ordering in the $z$-direction, the thermal averages $\braket{S^\Vert_{\boldsymbol{R}}} = \braket{S^\perp_{\boldsymbol{R}}} = 0$, and
we can write the mean field Hamiltonian as
$\mathcal{H} = -1/2 \sum_{\boldsymbol{r}} S^z_{\boldsymbol{r}}B^z_{\boldsymbol{r}}$
with
 $B^z_{\boldsymbol{r}}=\sum_{\boldsymbol{R}}[J_I\delta_{\boldsymbol{R},\boldsymbol{\gamma}}+J_{zz}(\boldsymbol{R})]\braket{S^z_{\boldsymbol{r}+\boldsymbol{R}}}$.
 For a ferromagnetic state where the magnetization $\braket{S^z_{\boldsymbol{R}}}$ is spatially independent, the $\sum_{\boldsymbol{R}}$
 reduces to $\small{\cal Z}J_I + J_{zz}(\boldsymbol{Q}=0)$ where ${\cal Z}$ is the lattice coordination number. It is instructive to write 
 $J_{zz}(\boldsymbol{Q}=0) = J_0 \chi_{VV}$ in terms of the Van Vleck (VV) susceptibility, as discussed at the end of the previous Section.
A standard calculation yields $T_c = T_{c0} + \delta T_c$, where $T_{c0}$ arises from the intrinsic $J_I$ while
the enhancement due to the BR interaction is
\begin{equation}\label{Eq:TcEnhancement}
    \delta T_c=\frac{J_0 S(S+1)}{3k_B}\chi_{VV}
\end{equation}
The VV susceptibility is considered to be the primary mechanism driving magnetic order in TI films doped with magnetic impurities, and is greatly enhanced by band inversion ~\cite{yu2010QAHE,Wang2015}.
Here, we find that it also plays a crucial role in enhancing $T_c$ in FMI/TI bilayers.

To estimate $\delta T_c$ we compute the exchange coupling $J_{zz}$, or equivalently $\chi_{VV}$, 
by setting the gap $\Delta =0$. This is because at the onset of ordering, i.e., just above $T_c$, there is no TR symmetry breaking.
However, we do not need to put in any other thermal effects in the exchange coupling expressions.
In Appendix \ref{APP:VVFiniteT}, we show by explicitly taking into account thermal factors that the
BR interactions are essentially temperature independent up to and beyond room temperature, so long as $k_B T \ll W$ (the bandwidth). Thus we simply use the zero-temperature expressions in \eqref{Eq:BRCouplings} with $\xi = R/R_\Delta = 0$ since $\Delta$ vanishes.


\section{Effect of TI Thickness on $T_c$ enhancement}\label{Sec.V}

Up to this point we have analyzed the problem of an essentially semi-infinite TI film.
We now address the question of how these results are impacted by the 
the thickness of the TI film. With decreasing thickness, the 2D surface states on the top and bottom layers of the TI
can hybridize, opening up a gap even in the absence of any TR symmetry breaking. (The exchange gap $\Delta$ is zero,
in our analysis of $T_c$, as explained above). Our goal is to compute how this hybridization, and topological transition
in the surface states, affects the BR interaction and $T_c$ enhancement.

We use the top-bottom basis 
$\psi(\boldsymbol{k})=(c_{\boldsymbol{k}\uparrow}^t, c_{\boldsymbol{k}\downarrow}^t, c_{\boldsymbol{k}\uparrow}^b, c_{\boldsymbol{k}\downarrow}^b)^T$ where $c_{\boldsymbol{k}\sigma}^{t(b)}$ annihilates an electron with momentum $\boldsymbol{k}=(k_x,k_y)$ and spin $\sigma$ on the top (bottom) surface of the TI.
These are described by the Hamiltonian~\cite{Zhang20106QL, Shan_2010, Wang2015}
\begin{equation}\label{Eq:H0R}
\begin{aligned}
        H_0=\tau_z\otimes [\hbar v_F (\boldsymbol{k}\times\boldsymbol{\sigma})\cdot\hat{z}]+m({\boldsymbol{k}})\tau_x\otimes\sigma_0
\end{aligned}
\end{equation}
where $\sigma_0$ and $\tau_0$ are identity matrices and $\sigma_i$, $\tau_i$ ($i=x,y,z$) are Pauli matrices operating in spin and surface subspaces, respectively. 
The last term $m(\boldsymbol{k})=m_0-m_1 k^2$ describes the hybridization gap between top and bottom surfaces,
which leads to a topological phase transition in the TI surface states. As shown in ref.~\cite{BHZ2006}, the TI thin film is a
quantum spin Hall (QSH) insulator for $m_0/m_1>0$ and a ``trivial" insulator for $m_0/m_1<0$.
Specifically, for (Bi,Sb)$_2$Te$_3$ (BST) thin films, this transition can be tuned by the film thickness~\cite{Liu2010Oscillatory}:
4 quintuple layer (QL) BST is a QSH insulator, whereas 3 QL BST is a ``trivial" insulator~\cite{Zhang20106QL, Wang2015}. 

We use Eq.~\eqref{Eq:H0R} to derive the static spin susceptibility that generalizes Eq.~\eqref{Eq:chi} taking into account the 
hybridization between top and bottom surfaces. We assume that the local moments in the FMI interact only with the top surface of the TI, and
thus $\chi^{tt}_{zz}(\boldsymbol{R})$ that involves the top electrons enters our analysis. For the definition of $\chi^{tt}_{zz}$ and its calculation, we refer the reader
to Appendix \ref{App:BRIntTIThinFilms}; here we only discuss the results.

\begin{figure}[t]
  \centering
  \includegraphics[width=0.47\linewidth]{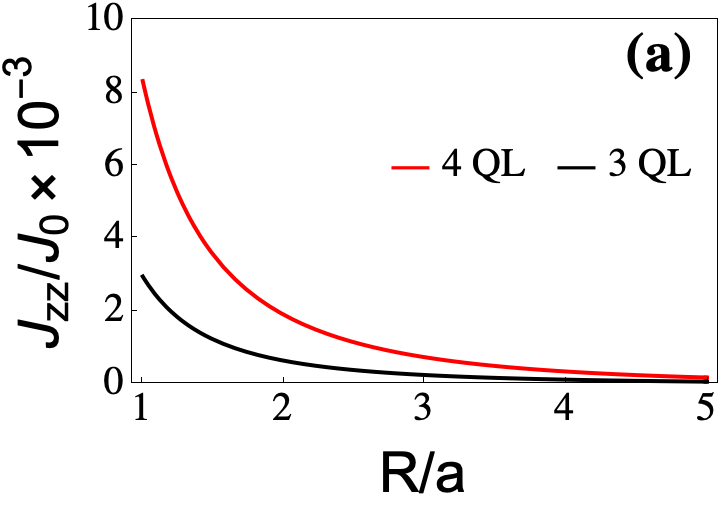}\hfill
  \includegraphics[width=0.49\linewidth]{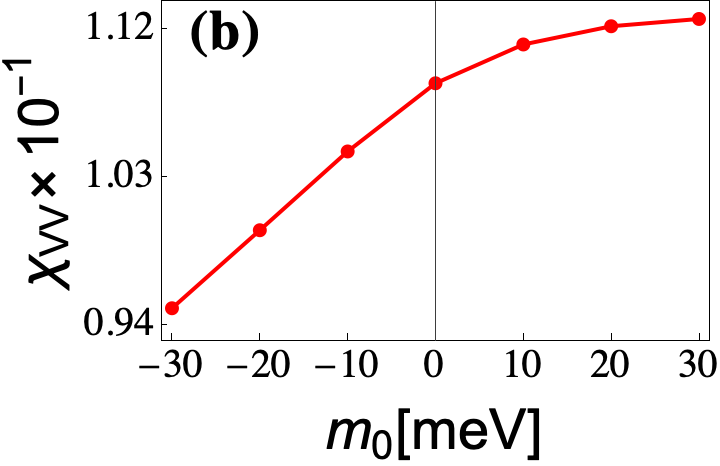}
  \caption{\justifying (a) $J^{tt}_{zz}(R)$ in units of $J_0$ between two local moments on the top surface for 4 QL of BST (red, $\hbar v_F=2.43$ eV$\cdot$\AA, $m_0=29$ meV, $m_1=12.9$ eV$\cdot$\AA$^2$) 
  and 3 QL of BST (black, $\hbar v_F=2.43$ eV$\cdot$\AA, $m_0=-44$ meV, $m_1=37.3$ eV$\cdot$\AA$^2$). 
  (b) Dependence of $\chi^{tt}_{zz}(\boldsymbol{q}=0)\equiv\chi^{tt}_{VV}$ on $m_0$, $\hbar v_F=2.43$ eV$\cdot$\AA, $m_1=12.9$ eV$\cdot$\AA$^2$. The black line at $m_0=0$ is the point of band inversion. 
   Parameters from ref.~\cite{Wang2015}.}
  \label{fig:BRThinFilm}
\end{figure}

Figure~\ref{fig:BRThinFilm} shows the BR interaction between two local magnetic moments on the top surface taking into account surface hybridization
using realistic parameters~\cite{Wang2015} for BST (noted in the Figure caption). We see from panel (a) that the BR interaction for 
$R \sim a$ in 4 QL BST is stronger than in 3 QL BST. 
 To understand this, we plot in panel (b), the VV susceptibility $\chi^{tt}_{VV}=\sum_{\boldsymbol{R}}\chi^{tt}_{zz}(\boldsymbol{R})$ 
 as a function of $m_0$, which controls the hybridization gap, with all other parameters kept constant.  
 We note that $\chi^{tt}_{VV}$ increases smoothly across the topological transition at $m_0 = 0$, 
 reminiscent of the enhancement across the band inversion point as a function of spin-orbit coupling reported in ref.~\cite{yu2010QAHE}. 
Comparing these thin film results to the results of the previous Section we see, however, that the hybridization between surfaces weakens
the strength of the BR interaction compared to what we found for the thick (semi-infinite) TI film.


\section{Comparison with experiments}\label{Sec.VI}

We finally turn to the comparison of our results for $T_c$ enhancement with experiments on one and two monolayers of Cr$_2$Te$_3$ proximitized by (Bi,Sb)$_2$Te$_3$~\cite{ou2023enhanced}. Atomically thin Cr$_2$Te$_3$ is known to be a FMI~\cite{shen2023}.
Each layer has two inequivalent Cr sites, but for simplicity we treat the Cr1 and Cr2 sites as identical and model the FMI
as a triangular lattice of Cr local moments. 
We use known parameters for this system: spin $S=3/2$, lattice constant $a=3.93$ \AA, Fermi velocity $v_F=3.69\times 10^5$ m/s~\cite{BST2015}, with the 
bandwidth given by $W=\hbar v_F/a$. We also need the coupling $J$ between the surface states and local moments (see eq.~\eqref{Eq:H}),
for which we use the estimate $J\approx 0.1$ eV from ref.~\cite{Zhang2009}. 

The Van Vleck susceptibility $\chi_{VV}$ is $T$-independent in the regime of interest $k_B T \ll W$ 
(see Appendix~\ref{APP:VVFiniteT}) and thus we can use the $T=0$ results derived above, except that we set the 
the TR symmetry breaking gap $\Delta = 0$ in our calculation of $T_c$.
Using the parameters given above, and using the results of Sec.~\ref{Sec.IV} for a
TI layer that is thick enough that we can ignore hybridization between surfaces,
we find that $\delta T_c\approx 50$ K.
On the other hand, when hybridization is important we need to use the results of Sec.~\ref{Sec.V} and Appendix \ref{App:BRIntTIThinFilms}
to estimate $\delta T_c$ for the FMI layer next to 4 QL and 3 QL BST.
We conclude that the $T_c$ enhancements are 
$$
\delta T_c({\rm 3 QL}):\delta T_c({\rm 4 QL}):\delta T_c({\rm Thick})  = 0.23: 0.63: 1
$$
Note that these ratios are independent of the coupling $J$ between the surface states and the local moments.

We note that the Cr$_2$Te$_3$/(Bi,Sb)$_2$Te$_3$ experiments of ref.~\cite{ou2023enhanced} were on samples with 4QLs of BST,
for which we find a $\delta T_c \approx 30$ K.
This is in reasonable agreement with the experimental value of $\delta T_c = 20$ K~\cite{ou2023enhanced},
given the simplicity of our modeling with mean field theory which overestimates $T_c$
and the uncertainties in estimating $J$.

\section{Conclusions}\label{Sec.VII}

To summarize our main conclusions, we have investigated how interactions mediated by the surface states of a topological insulator (TI) influence the magnetism in an adjacent two-dimensional ferromagnetic insulator (FMI). We model the TI surface states as Dirac electrons that may 
acquire a mass term induced by magnetic ordering. 
Since the magnetic moments are arranged on the FMI lattice, our analysis focuses on effective interactions on the scale of the lattice constant. The chemical potential lies either within the gap or corresponds to a small Fermi wave vector $k_F$.
Consequently, the regime of interest here differs substantially from previous studies that primarily considered dilute magnetic impurities on TI surfaces.

We show that in this regime, the Bloembergen-Rowland (BR) interaction arising from virtual excitations between the valence and conduction bands dominates over the more familiar oscillatory Ruderman- Kittel- Kasuya- Yosida (RKKY) interaction, which originates from the Fermi surface. Due to the spin–momentum locking of TI surface states, the BR interaction exhibits a characteristic anisotropy that favors out-of-plane magnetic ordering in the FMI. Our mean-field analysis further shows that the Curie temperature $T_c$ of the FMI is enhanced by an amount proportional to the Van Vleck susceptibility of the TI surface electrons. We also analyze the effect of hybridization between the top and bottom surfaces of a thin TI film and provide a quantitative comparison with recent experiments on
Cr$_2$Te$_3$/(Bi,Sb)$_2$Te$_3$ heterostructures.

In summary, our analysis highlights the interplay between topological surface states of a TI and magnetism in FMI-TI heterostructures and gives insight into the experimentally observed enhancement of the FM $T_c$. These findings pave the way for further exploration and optimization of such heterostructures for potential applications in spintronics and quantum technologies.

\section{Acknowledgments}

We thank Jagadeesh Moodera and Hang Chi for bringing this problem to our attention and for very useful discussions about their experiments.
We acknowledge support from the NSF Materials Research Science and Engineering Center Grant No.~DMR-2011876.

\section{Appendices} 
\appendix

\section{Surface State Mediated interaction between Moments}\label{App:RKKY}

In this appendix, we compute the exchange between local moments mediated by the surface states, as defined in Eqs.~\eqref{Eq:Hex}-\eqref{Eq:chi}. In the weak-coupling limit $J\ll W$ (with $W$ the electronic bandwidth), the interaction $H_{\text{int}}$ is treated perturbatively in relation to the non-interacting Hamiltonian $H_0(\boldsymbol{k})$. The free energy of $H=H_0+H_{\text{int}}$ is
\begin{equation}\label{Eq:FreeEnergy}
\begin{aligned}
F &= -\frac{1}{\beta}\ln \operatorname{Tr}\, e^{-\beta(H_0+\lambda H_{\text{int}})} \\
  &= F_0 - \frac{1}{\beta}\ln \langle \hat{U}(\beta,0)\rangle_0 ,
\end{aligned}
\end{equation}
where $\beta = 1/k_B T$, $F_0$ is the free energy of $H_0$, $\lambda$ is a bookkeeping parameter for the perturbation series, and $\langle \hat{U}(\beta,0)\rangle_0$ is the expectation value of the time-ordered exponential \cite{AGD1965Pergamon, Coleman2015}, 
\begin{equation} 
\begin{aligned} 
\langle \hat{U} (\beta,0)\rangle_0 = 1 &+ \sum_{n=1}^{\infty}\frac{(-\lambda)^n}{n!}\int_0^\beta d\tau_1\cdots \int_0^\beta d\tau_n \\ &\quad\times \langle T_\tau [H_{\text{int}}(\tau_1)\cdots H_{\text{int}}(\tau_n)]\rangle_0  
\end{aligned}
\end{equation}
Here, $T_\tau$ denotes imaginary-time ordering and $H_{\text{int}}(\tau)=e^{\tau H_0}H_{\text{int}}e^{-\tau H_0}$ is the interaction-picture operator.

Expanding the logarithm in Eq.~\eqref{Eq:FreeEnergy} around unity up to $\mathcal{O}(\lambda^3)$ gives
\begin{equation}
    F = F_0 + \frac{\lambda}{\beta}e_1 - \frac{\lambda^2}{2\beta}(e_2-e_1^2) + \mathcal{O}(\lambda^3),
\end{equation}
with
\begin{equation}
\begin{aligned}
    e_1 &= \int_0^\beta d\tau \,\langle H_{\text{int}}(\tau)\rangle_0 , \\
    e_2 &= \int_0^\beta d\tau_1 \int_0^\beta d\tau_2 \,\langle T_\tau[H_{\text{int}}(\tau_1) H_{\text{int}}(\tau_2)]\rangle_0 .
\end{aligned}
\end{equation}

For the interaction Hamiltonian
\begin{equation}
    H_{\text{int}} = J a^2 \sum_{\boldsymbol{r}} S_i(\boldsymbol{r})\, c^\dagger_\alpha(\boldsymbol{r}) \sigma^i_{\alpha\beta}\, c_\beta(\boldsymbol{r}),
\end{equation}
where repeated indices are summed, the first-order term becomes
\begin{equation}
\begin{aligned}
    e_1 &= J a^2 \sum_{\boldsymbol{r}} S_i(\boldsymbol{r}) \int_0^\beta d\tau\, 
           \langle c^\dagger_\alpha(\boldsymbol{r},\tau)\sigma^i_{\alpha\beta} c_\beta(\boldsymbol{r},\tau)\rangle_0 \\
        &= J\beta \sum_{\boldsymbol{r}} S_i(\boldsymbol{r})\, \mathrm{Tr}\big[\sigma^i G(\boldsymbol{r}=0,\tau=0^-)\big]
\end{aligned}
\end{equation}
where $G(\boldsymbol{r},\tau)$ is the free-fermion real-space and imaginary-time Green's function \cite{AGD1965Pergamon}, with $a^2$ absorbed into its normalization. Fourier transforming this and using $G(\boldsymbol{r},\tau)=G_\nu(\boldsymbol{r},\tau)\sigma^\nu$, we obtain
\begin{equation}
  e_1= J\beta \sum_{\boldsymbol r} S_i(\boldsymbol r)
    \int \frac{d^2\boldsymbol k}{(2\pi)^2}\,
    \mathrm{Tr}\big[\sigma^i \sigma^\nu\big]\,
    G_\nu(\boldsymbol k,\tau=0^-)
\end{equation}
Here $\nu\in\{0,1,2,3\}$. The $\nu=0$ term vanishes under the trace, and the $\nu=1,2$ terms vanish upon $\boldsymbol k$-integration [since $G_{1,2}(\boldsymbol k)$ are odd in $\boldsymbol k$; see Appendix~\ref{App:RealSpaceGF}]. The $\nu=3$ term can be finite if time-reversal  symmetry is broken, either by an external magnetic field or by spontaneous magnetic order. In this work we do not apply an external magnetic field, and any contribution from spontaneously broken TR symmetry would arise at higher order in $J$ and is therefore neglected.

The second-order contribution is
\begin{equation}
\begin{aligned}
    & \qquad e_2 = J^2 a^4 \sum_{\boldsymbol{x},\boldsymbol{y}} S_i(\boldsymbol{x}) S_j(\boldsymbol{y}) 
            \int_0^\beta d\tau_1 \int_0^\beta d\tau_2 \\
        &\times \langle T_\tau [c^\dagger_\alpha(\boldsymbol{x},\tau_1)\sigma^i_{\alpha\beta} c_\beta(\boldsymbol{x},\tau_1)
            c^\dagger_m(\boldsymbol{y},\tau_2)\sigma^j_{mn} c_n(\boldsymbol{y},\tau_2)]\rangle_0 
\end{aligned}
\end{equation}
Using Wick's theorem and fermionic antiperiodicity \cite{AGD1965Pergamon}, this reduces to
\begin{equation}
\begin{aligned}
    e_2 &= - J^2 \beta \sum_{\boldsymbol{x},\boldsymbol{y}} S_i(\boldsymbol{x}) S_j(\boldsymbol{y}) 
            \int_0^\beta d\tau\, \\
        &\qquad\times \mathrm{Tr}\!\left[\sigma_i G(\boldsymbol{x}-\boldsymbol{y},\tau)\,
                                \sigma_j G(\boldsymbol{y}-\boldsymbol{x},-\tau)\right].
\end{aligned}
\end{equation}

Thus, the free energy up to order $J^2$ is
\begin{equation}
\begin{aligned}
    F &= F_0
       - \frac{J^2}{2W}\sum_{\boldsymbol{x},\boldsymbol{y}}
             S_i(\boldsymbol{x})\, \chi_{ij}(\boldsymbol{x}-\boldsymbol{y})\, S_j(\boldsymbol{y}) 
             + \mathcal{O}(\frac{J^3}{W^2})
\end{aligned}
\end{equation}
where the factor $1/2$ compensates for double counting, so the surface state-mediated exchange Hamiltonian for two local moments is given as 

\begin{equation}
\begin{aligned}
        \mathcal{H}_{\rm ex}=-{J^2\over W}\sum_{ij}\chi_{ij}(\boldsymbol{R}_1-\boldsymbol{R}_2)S^1_i S^2_j;
\end{aligned}
\end{equation}
with $(i,j)=(x,y,z)$ as expressed in Eq.\eqref{Eq:Exchange} in the main text, with the spin susceptibility

\begin{equation}\label{Eq:SusceptibilityTau}
    \chi_{ij}(\boldsymbol{R}) = -W\int_0^\beta d\tau\, 
      \mathrm{Tr}\!\left[\sigma_i G(\boldsymbol{R},\tau)\, \sigma_j G(-\boldsymbol{R},-\tau)\right].
\end{equation}
where we made it dimensionless. Expressing the Green's functions in Matsubara frequencies \cite{AGD1965Pergamon},
\begin{equation}
    G(\tau) = \frac{1}{\beta}\sum_{i\omega_n} e^{-i\omega_n \tau} G(i\omega_n),
\end{equation}
we obtain the static susceptibility as
\begin{equation}\label{Eq:SusceptibilityMatsubara}
    \chi_{ij}(\boldsymbol{R}) = -\frac{W}{\beta}\sum_{i\omega_n} 
        \mathrm{Tr}\!\left[\sigma_i G(\boldsymbol{R},i\omega_n)\,
                          \sigma_j G(-\boldsymbol{R},i\omega_n)\right].
\end{equation}
Finally, performing the Matsubara sum using the poles of the Fermi function 
$n_F(\omega) = [e^{\beta(\omega-\mu)}+1]^{-1}$ yields
\begin{equation}
\begin{aligned}
    \chi_{ij}(\boldsymbol{R}) &= \frac{W}{\pi}\,\mathrm{Im}\!
        \int_{-\infty}^\infty d\omega\, n_F(\omega)\,\\
        &\times\mathrm{Tr}\!\Big[\sigma_i G(\boldsymbol{R},\omega^+)\,
                          \sigma_j G(-\boldsymbol{R},\omega^+)\Big],
\end{aligned}
\end{equation}
from which the zero-temperature expression for $\chi_{ij}(\boldsymbol{R})$ as expressed in Eq.\eqref{Eq:chi} follows.

We note that in Eq.~\eqref{Eq:SusceptibilityMatsubara}, two equivalent procedures can in principle be employed. 
The first approach evaluates the real-space Green's function and subsequently performs the Matsubara frequency sum, as done above. 
Alternatively, one may first carry out the Matsubara summation in momentum space and then perform the momentum integration \cite{Saremi2007, Garate2010RKKY}. 
In the latter case, the real-space susceptibility can be written as
\begin{equation}
    \chi_{ij}(\boldsymbol{R})
    = \int \frac{d^2\boldsymbol{q}}{(2\pi/a)^2}\,
      e^{i\boldsymbol{q}\cdot\boldsymbol{R}}\,
      \chi_{ij}(\boldsymbol{q}),
\end{equation}
where the momentum-space static susceptibility takes the form
\begin{equation}
\begin{aligned}
    \chi_{ij}(\boldsymbol{q})
    = -&\frac{W}{\beta}
      \sum_{i\omega_n}
      \int \frac{d^2\boldsymbol{k}}{(2\pi/a)^2}\\
      &\times\mathrm{Tr}\!\left[
        \sigma_i G(\boldsymbol{k}+\boldsymbol{q},i\omega_n)\,
        \sigma_j G(\boldsymbol{k},i\omega_n)
      \right],
\end{aligned}
\end{equation}
Switching the order of the Matsubara summation and the momentum integration yields integrals of the form
\begin{equation}
\begin{aligned}
    \chi_{ij}(\boldsymbol{q})
    &= \int \frac{d^2\boldsymbol{k}}{(2\pi/a)^2}
       \sum_{mn}
       \Lambda^{mn}_{ij}(\boldsymbol{k},\boldsymbol{q}) \\
    &\quad \times
       \frac{n_F[\epsilon_m(\boldsymbol{k}+\boldsymbol{q})]
       - n_F[\epsilon_n(\boldsymbol{k})]}
       {\epsilon_n(\boldsymbol{k})
       - \epsilon_m(\boldsymbol{k}+\boldsymbol{q})},
\end{aligned}
\end{equation}
where $\Lambda_{ij}^{mn}(\boldsymbol{k},\boldsymbol{q})$ is a matrix depending on the components $(i,j)$ and band indices $(m,n)$; its explicit form is not essential here, and $\epsilon_n(\boldsymbol{k})$ denotes the band energy. Within this formalism, it becomes evident that even at $\mu=0$, the susceptibility can remain finite due to virtual interband transitions, which is known as the Bloembergen-Rowland contribution \cite{BR1955}. For Dirac electrons, however, because the energy dispersion is linear in $\boldsymbol{k}$, the integral exhibits an ultraviolet divergence and therefore requires the introduction of a high-momentum cutoff \cite{Saremi2007}. Several studies have addressed this issue using cutoff-based regularization schemes and obtained results consistent with those derived from the real-space Green's function approach \cite{Saremi2007,  Zhang2009, Garate2010RKKY, Sau2016}. In this work, we adopt the latter approach to ensure well-defined convergence without introducing an explicit cutoff.

\section{Real-space Green's function for Dirac electrons}\label{App:RealSpaceGF}

Here, we compute the real-space Green's function $G(\boldsymbol{R},\omega^+)$ required for the exchange couplings in Eq.~\eqref{Eq:Exchange} following ref. \cite{dugaev1994GFs}. The momentum-space (retarded) Green's function at frequency $\omega$ is \cite{AGD1965Pergamon}
\begin{equation}\label{GF}
\begin{aligned}
G(\boldsymbol{k},\omega^+)
&= \frac{1}{\omega^+ - H_0(\boldsymbol{k})} \\
&= \frac{\omega^+\sigma_0 + \hbar v_F(\boldsymbol{k}\!\times\!\boldsymbol{\sigma})_z + \Delta\sigma_z}
        {(\omega^+)^2 - \hbar^2 v_F^2 k^2 - \Delta^2},
\end{aligned}
\end{equation}
with $H_0(\boldsymbol{k})=\hbar v_F(\boldsymbol{k}\!\times\!\boldsymbol{\sigma})_z+\Delta\sigma_z$ and $\omega^+=\omega+i0^+$. The real-space Green's function follows by Fourier transformation,
\begin{equation}
G(\boldsymbol{R},\omega^+)=\int \frac{d^2\boldsymbol{k}}{(2\pi/a)^2}\,
e^{i\boldsymbol{k}\cdot\boldsymbol{R}}\, G(\boldsymbol{k},\omega^+).
\end{equation}
This requires integrals of the type
\begin{equation}
\begin{aligned}
    Q_0(\boldsymbol{R}) &= \int \frac{d^2\boldsymbol{k}}{(2\pi)^2}\,
       e^{i\boldsymbol{k}\cdot\boldsymbol{R}}\frac{1}{A^2-k^2}, \\
    \boldsymbol{Q}_1(\boldsymbol{R}) &= \int \frac{d^2\boldsymbol{k}}{(2\pi)^2}\,e^{i\boldsymbol{k}\cdot\boldsymbol{R}}
       \frac{\boldsymbol{k}\,}{A^2-k^2}.
\end{aligned}
\end{equation}
Switching to polar coordinates, the angular integration yields
\begin{equation}
    Q_0(\boldsymbol{R}) = \frac{1}{2\pi}\int_0^\infty dk \,\frac{k J_0(kR)}{A^2-k^2},
\end{equation}
where $J_n$ is the Bessel function of order $n$. Using $J_n(z) = \tfrac{1}{2}[\text{H}_n^{(1)}(z)+\text{H}_n^{(2)}(z)]$ and deforming the contour just above the real axis, one finds
\begin{equation}\label{Eq:A7}
    Q_0(\boldsymbol{R}) = \frac{1}{4\pi}\int_{-\infty+i\delta}^{\infty+i\delta} 
        dk \,\frac{k\, \text{H}_0^{(1)}(kR)}{A^2-k^2}.
\end{equation}
Closing the contour in the upper half-plane and applying the residue theorem gives
\begin{equation}
    Q_0(\boldsymbol{R}) = -\frac{i}{4}\, \text{H}_0^{(1)}(AR),
\end{equation}
where $A = \pm i\sqrt{[\Delta^2-(\omega^+)^2]}/\hbar v_F$ and the root is chosen with $\mathrm{Im}(A)>0$.

For the vector integral,
\begin{equation}
\begin{aligned}
    \boldsymbol{Q}_1(\boldsymbol{R}) 
      &= -i\,\frac{\partial Q_0(\boldsymbol{R})}{\partial \boldsymbol{R}} \\
      &= \frac{A}{4R}\,\text{H}_1^{(1)}(AR)\,\boldsymbol{R},
\end{aligned}
\end{equation}
with $\text{H}_1^{(1)}$ the Hankel function of order one.

Collecting terms, the real-space Green's function is
\begin{equation}\label{Eq:GFs}
\begin{aligned}
    G(\boldsymbol{R},\omega^+)
       &= -\frac{i a^2}{4\hbar^2 v_F^2}
          (\omega^+\sigma_0+\Delta\sigma_z)\,
          \text{H}_0^{(1)}\!\left(\tfrac{iR\mathcal{E}}{\hbar v_F}\right) \\
       &\quad + \frac{i a^2 \mathcal{E}}{4\hbar^2 v_F^2 R}\,
          [\boldsymbol{R}\times\boldsymbol{\sigma}]_z\,
          \text{H}_1^{(1)}\!\left(\tfrac{iR\mathcal{E}}{\hbar v_F}\right),
\end{aligned}
\end{equation}
where $\mathcal{E}=\sqrt{\Delta^2-(\omega^+)^2}$, as introduced in the main text.

\section{Exchange Couplings for $\mu\neq0$ and $\Delta=0$ }\label{App:nonzeromu}

Here, we present the analytical expressions for the integrals in Eq.\eqref{Eq:Exchange} when $\mu \neq 0$ and $\Delta = 0$. We need to evaluate integrals of the form
\begin{equation}
\begin{aligned}
I=\text{Im}\int_{-\infty}^\mu d\omega (\omega^+)^2\text{H}_n^{(1)}(iR\sqrt{-(\omega^+)^2}/\hbar v_F)^2
\end{aligned}
\end{equation}
where we can split the integral into two pieces: $\int_{-\infty}^0+\int_0^\mu$. The first piece gives \cite{gradshteyn2014table, Yarmohammadi2023}

\begin{equation}
    \begin{aligned}
        &\text{Im}\int_{-\infty}^0 d\omega (\omega^+)^2\text{H}_n^{(1)}(iR\sqrt{-(\omega^+)^2}/\hbar v_F)^2=\\
        &\int_{0}^\infty dy y^2\text{H}_n^{(1)}(iRy/\hbar v_F)^2= (4n-1)\frac{\hbar^3v_F^3}{8R^3}, \quad n=0,1
    \end{aligned}
\end{equation}
where we used contour integral in the second quadrant and the fact that $\text{H}_n^{(1)}(z) \rightarrow 0$ as $|z|\rightarrow \infty$. 

The second piece can be written in terms of the special Meijer-G function $\mathbb{G}_{m,n}^{p,q}$ \cite{gradshteyn2014table}

\begin{equation}
\begin{aligned}
&\text{Im}\int_{0}^\mu d\omega (\omega^+)^2\text{H}_n^{(1)}(iR\sqrt{-(\omega^+)^2}/\hbar v_F)^2=\\
&-\frac{\mu^3}{\sqrt{\pi}}\mathbb{G}_{2,4}^{2,1}\left(
  \begin{matrix}
  &  -\frac{1}{2} & \frac{1}{2} \\
  0 & n & -\frac{3}{2} & -n 
  \end{matrix}\Big|
  \begin{matrix}
   (k_F R)^2
  \end{matrix}
\right)
\end{aligned}
\end{equation}
where $k_F=\mu/\hbar v_F$. Using these, Eq.\eqref{Eq:Exchange} reduces to
\begin{equation}
    \begin{aligned}
        J_{zz}(\boldsymbol{R})&=\frac{J_K^2}{16\pi\hbar v_F R^3}\Big(1+\frac{2}{\sqrt{\pi}}(k_FR)^3\\
        &\times\Big[\mathbb{G}_{2,4}^{2,1}\left(
  \begin{matrix}
  &  -\frac{1}{2} & \frac{1}{2} \\
  0 & 1 & -\frac{3}{2} & -1 
  \end{matrix}
\Big|
  \begin{matrix}
    (k_FR)^2
  \end{matrix}
\right)\\
&-\mathbb{G}_{2,4}^{2,1}\left(
  \begin{matrix}
  &  -\frac{1}{2} & \frac{1}{2} \\
  0 & 0 & -\frac{3}{2} & 0 
  \end{matrix}
\Big|
  \begin{matrix}
    (k_FR)^2
  \end{matrix}
\right)\Big]\Big)\\
        J_{\perp}(\boldsymbol{R}) &=-\frac{J_K^2}{32\pi\hbar v_F R^3}\Big(1+\frac{4}{\sqrt{\pi}}(k_FR)^3\\
        &\times\Big[\mathbb{G}_{2,4}^{2,1}\left(
  \begin{matrix}
  &  -\frac{1}{2} & \frac{1}{2} \\
  0 & 1 & -\frac{3}{2} & -1 
  \end{matrix}
\Big|
  \begin{matrix}
   (k_FR)^2
  \end{matrix}
\right)\\
&+\mathbb{G}_{2,4}^{2,1}\left(
  \begin{matrix}
  &  -\frac{1}{2} & \frac{1}{2} \\
  0 & 0 & -\frac{3}{2} & 0 
  \end{matrix}
  \Big|
  \begin{matrix}
   (k_FR)^2
  \end{matrix}
\right)\Big]\Big)\\
J_{DM}(\boldsymbol{R}) &=-\frac{J_K^2}{4\pi^{3/2}\hbar v_F R^3}\Big(\Big[\mathbb{G}_{2,4}^{2,1}\left(
  \begin{matrix}
  &  2 & \frac{3}{2} \\
  2 & 2 & 0 & 1 
  \end{matrix}
\Big|
  \begin{matrix}
   (k_FR)^2
  \end{matrix}
\right)\\
&-\mathbb{G}_{1,3}^{2,0}\left(
  \begin{matrix}
  &   \frac{3}{2} \\
  1 & 2 & 0  
  \end{matrix}
\Big|
  \begin{matrix}
   (k_FR)^2
  \end{matrix}
\right)\Big]\Big)
    \end{aligned}
\end{equation}
By expanding the Meijer-G functions near the origin and as $k_FR\rightarrow \infty$, we derive the behavior of the exchange couplings as presented in Eq.\eqref{Eq:smallkFR} and their long-range oscillatory nature, $R^{-2}\cos{(2k_FR+\phi)}$, respectively.

\section{Ground State and Excitation Spectrum of BR-Coupled Lattice of Spins}\label{App:BRGroundState}

In this appendix, we determine the magnetic ordering of the BR-coupled lattice by (i) minimizing the few-spin energetics and (ii) computing the spin-wave spectrum of the full Hamiltonian in Eq.~\eqref{Eq:BRHamiltonian}.

\textit{Two spins.} For a nearest-neighbor bond $R=a\ll R_\Delta$,
\begin{equation}
\mathcal{H}_{\rm ex}
=-\tilde{J}_0 S_1^z S_2^z+\frac{\tilde{J}_0}{2} S_1^\perp S_2^\perp - \tilde{J}_0 S_1^{\parallel} S_2^{\parallel}
\end{equation}
where $\tilde{J}_0=J_0/16\pi$ is the exchange constant and $\parallel$ and $\perp$ are defined relative to $\hat{\boldsymbol{R}}$. Minimization gives two degenerate ferromagnetic minima with energy $-\tilde{J}_0S^2$: (i) out-of-plane (OOP) alignment $(\theta_1,\theta_2)=(0,0)$ or $(\pi,\pi)$, and (ii) in-plane (IP) alignment along the bond, $(\theta_1,\theta_2)=(\tfrac{\pi}{2},\tfrac{\pi}{2})$ with $(\phi_1,\phi_2)=(\theta_R,\theta_R)$ or $(\theta_R{+}\pi,\theta_R{+}\pi)$ where $\theta_R=\tan^{-1}(R_y/R_x)$. Antiferromagnetic alignment is maximal in energy, $+\tilde{J}_0S^2$.

\textit{Beyond two spins.} The IP state depends on bond direction and becomes frustrated once additional, non-collinear bonds are present. For three spins on an equilateral triangle of side $a$, the ground state is OOP ferromagnetic with energy $-3\tilde{J}_0S^2$, whereas the best coplanar IP configuration at $(\theta_1,\theta_2,\theta_3)=(\tfrac{\pi}{2},\tfrac{\pi}{2},\tfrac{\pi}{2})$ reaches only $-\tfrac{15}{8}\tilde{J}_0S^2$. Hence, in 2D the BR coupling selects an OOP ferromagnet.

\begin{figure}[t]
  \centering
  \includegraphics[width=0.5\linewidth]{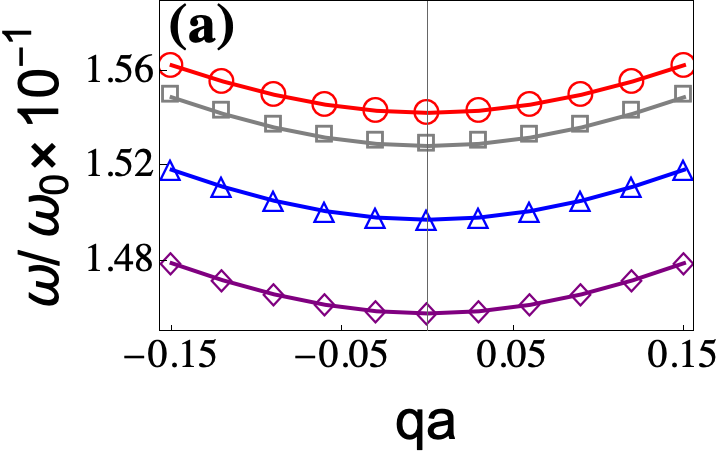}\hfill
  \includegraphics[width=0.5\linewidth]{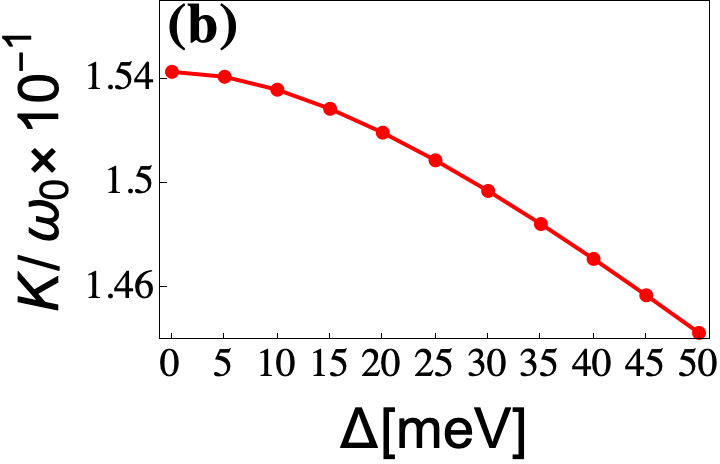}
  \caption{\justifying (a) Spin-wave spectrum $\omega/\omega_0$ (with $\omega_0 = S J_0$) as a function of $qa$ near $\boldsymbol{q}=0$ for different gap values:
$\Delta = 0~\text{meV}$ (red circles), 
$\Delta = 10~\text{meV}$ (gray squares),
$\Delta = 20~\text{meV}$ (blue triangles), and 
$\Delta = 30~\text{meV}$ (purple diamonds). 
The dispersion follows $\omega(q) \!\approx\! K + A q^2$ with $K = f(\Delta/W)\,\omega_0$. 
(b) The dimensionless parameter $f(\Delta/W) = K/\omega_0$ as a function of $\Delta$.}
  \label{fig:SWDispersion}
\end{figure}

\textit{Spin waves.} Expanding about the uniform OOP state ($S_{x,y}\ll S$) and using the Landau-Lifshitz equation \cite{landau1935theory},
\begin{equation}
\frac{\partial \boldsymbol{S}(\boldsymbol{r},t)}{\partial t}
=\boldsymbol{S}(\boldsymbol{r},t)\times\!\left[-\frac{\delta \mathcal{H}_{\rm ex}}{\delta \boldsymbol{S}(\boldsymbol{r},t)}\right],
\end{equation}
with the plane-wave ansatz $S_x=A\,e^{i(\boldsymbol{q}\cdot\boldsymbol{r}-\omega t)}$, $S_y=B\,e^{i(\boldsymbol{q}\cdot\boldsymbol{r}-\omega t)}$, $S_z\simeq S$, the linearized spectrum is
\begin{equation}\label{Eq:SWspectrum}
\omega(\boldsymbol{q})
=\sqrt{\big[J_{zz}(0)-J_1(\boldsymbol{q})\big]\big[J_{zz}(0)-J_2(\boldsymbol{q})\big]-J_{3}^2(\boldsymbol{q})},
\end{equation}
where $J_j(\boldsymbol{q})=\sum_{\boldsymbol{R}}e^{i\boldsymbol{q}\cdot\boldsymbol{R}}J_j(\boldsymbol{R})$ and $J_{zz}(0)\equiv J_{zz}(\boldsymbol{q}{=}0)$. The real-space couplings are
\begin{equation}\label{Eq:SWcouplings}
\begin{aligned}
J_1(\boldsymbol{R}) &= J_{\parallel}(\boldsymbol{R})-\!\big[J_{\parallel}(\boldsymbol{R})-J_{\perp}(\boldsymbol{R})\big]\cos^2\theta_R,\\
J_2(\boldsymbol{R}) &= J_{\perp}(\boldsymbol{R})+\!\big[J_{\parallel}(\boldsymbol{R})-J_{\perp}(\boldsymbol{R})\big]\cos^2\theta_R,\\
J_{3}(\boldsymbol{R}) &= \big[J_{\parallel}(\boldsymbol{R})-J_{\perp}(\boldsymbol{R})\big]\sin\theta_R\cos\theta_R,
\end{aligned}
\end{equation}
Equations~\eqref{Eq:SWspectrum} and \eqref{Eq:SWcouplings} make explicit the directional anisotropy of the couplings. Evaluating $\omega(\boldsymbol{q})$ across the Brillouin zone for various exchange gaps $\Delta$, we find a fully gapped dispersion, i.e. no mode softening or instabilities; and near $\boldsymbol{q}=0$ the spectrum is quadratic with a gap fixed by the Ising anisotropy [Fig.~\ref{fig:SWDispersion}]. This anisotropy stems from anisotropic nature of exchange couplings in Eq.\eqref{Eq:BRHamiltonian}

\section{BR Interaction for Thin TI films with Surface State Hybridization}\label{App:BRIntTIThinFilms}

Our goal here is to derive the static susceptibility  $\chi^{tt}_{zz}(\boldsymbol{R})$ of electrons on the top surface of a thin TI film taking into account hybridization effects with the bottom surface.  The static spin susceptibility at T=0 including the effect of surface hybridization effect is given as \cite{Coleman2015, Shiranzaei2017}

\begin{equation}\label{Eq:chiThinTI}
    \begin{aligned}
        \chi_{\alpha\beta}^{\nu\nu'}= &\frac{W}{\pi}\int_{-\infty}^0 d\omega\\
        & \times \text{Tr}[\sigma_\alpha G_{\nu\nu'}(\boldsymbol{R},\omega^+)\sigma_\beta G_{\nu'\nu}(-\boldsymbol{R},\omega^+)]
    \end{aligned}
\end{equation}
with $(\alpha,\beta)=(x,y,z)$ the spin components and $(\nu,\nu')=(t,b)$ denoting top and bottom surfaces. Here, we set $\mu=0$ for BR interaction. The momentum space retarded Green's function can be calculated as $G(\boldsymbol{k},\omega^+)=[\omega^+\tau_0\otimes\sigma_0-\tau_z\otimes \hbar v_F (\boldsymbol{k}\times\boldsymbol{\sigma})\cdot\hat{z}-m_{\boldsymbol{k}}\tau_x\otimes\sigma_0]^{-1}$, from which we obtain

\begin{equation}
    G(\boldsymbol{k},\omega^+)=\frac{\omega^++\hbar v_F(k_y\tau_z\sigma_x-k_x\tau_z\sigma_y) +m_{\boldsymbol{k}}\tau_x\sigma_0}{(\omega^+)^2-\epsilon_{\boldsymbol{k}}^2}
\end{equation}
where $\epsilon_{\boldsymbol{k}}=\pm\sqrt{\hbar^2v_F^2k^2+m_{\boldsymbol{k}}^2}$ and $m_{\boldsymbol{k}}=m_0-m_1k^2$. For two moments placed on the top surface, we are only interested in $G_{tt}(\boldsymbol{k},\omega^+)$. To obtain the real space retarded Green's function $G_{tt}(\boldsymbol{R},\omega^+)$, we need to evaluate integrals of the form

\begin{equation}
\begin{aligned}
    Q_0(\boldsymbol{R},\omega^+) &=\int \frac{d^2\boldsymbol{k}}{(2\pi/a)^2}e^{i\boldsymbol{k}\cdot\boldsymbol{R}}\frac{\omega^+}{(\omega^+)^2-\epsilon_{\boldsymbol{k}}^2}\\
    \boldsymbol{Q}_1(\boldsymbol{R},\omega^+) &=\int \frac{d^2\boldsymbol{k}}{(2\pi/a)^2}e^{i\boldsymbol{k}\cdot\boldsymbol{R}}\frac{\hbar v_F\boldsymbol{k}}{(\omega^+)^2-\epsilon_{\boldsymbol{k}}^2}
\end{aligned}
\end{equation}

Similarly to Appendix \ref{App:RealSpaceGF}, we obtain

\begin{equation}
\begin{aligned}
    Q_0(\boldsymbol{R},\omega^+) &=\frac{i a^2 \omega^+}{8 |m_1| \sqrt{(\omega^+)^2+\Delta_0^2/4}}\times\\
    &\times \Big[\text{H}_0^{(1)}(k_1R)-\text{H}_0^{(1)}(k_2R)\Big]\\
    \boldsymbol{Q}_1(\boldsymbol{R},\omega^+) &=\hat{\boldsymbol{R}}\frac{\hbar v_F a^2}{8 |m_1| \sqrt{(\omega^+)^2+\Delta_0^2/4}}\times\\
    &\times \Big[k_2\text{H}_1^{(1)}(k_2R)-k_1\text{H}_1^{(1)}(k_1R)\Big]
\end{aligned}
\end{equation}
where $k_{1,2}=i\sqrt{k_0^2\mp \frac{1}{|m_1|}\sqrt{(\omega^+)^2+\Delta_0^2/4}}$ with $k_0=\sqrt{(\hbar^2v_F^2-2m_0m_1)/2m_1^2}$ and $\Delta_0=\hbar v_F \sqrt{\hbar^2v_F^2-4m_0m_1}/|m_1| $. Here, we assumed $m_1\neq 0$ and we work in the regime where $m_0m_1<\hbar^2v_F^2/4$ such that both $k_0$ and $\Delta_0$ are real parameters. When $m_0m_1 > \hbar^2v_F^2/4$, the BR interaction is demonstrated to be oscillatory \cite{LitvinovBROscillatory}. We also made sure to select poles $k_{1,2}$ that are located in the upper half-plane for all frequencies $\omega^+$. 

Then, from Eq.\eqref{Eq:chiThinTI} we obtain the following.

\begin{equation}
    \chi_{zz}^{tt}(\boldsymbol{R})= \frac{2W}{\pi}\int_{-\infty+i0^+}^{i0^+} d\omega\big[|Q_0(\boldsymbol{R},\omega)|^2+|\boldsymbol{Q}_1(\boldsymbol{R},\omega)|^2\big]
\end{equation}
After analytically continuing $\omega+i0^+\to i\tilde{\omega}$, with the poles $k_{1,2}$ kept in the upper half-plane, this becomes

\begin{equation}
    \chi_{zz}^{tt}(\boldsymbol{R})= \frac{(W/\Delta_0)}{16\pi (R_0/a)^4}F(\boldsymbol{R}/R_0)
\end{equation}
where the dimensionless range function $F(\boldsymbol{R}/R_0)=F_0(\boldsymbol{R}/R_0)+F_1(\boldsymbol{R}/R_0)$ is given as 

\begin{equation}
    \begin{aligned}
        &F_0(\boldsymbol{R}/R_0) =\text{Im} \int_0^1 i \frac{dz}{z\sqrt{1-z^2}}\Big\{(z^2-1)\times\\
        & \Big[\text{H}_0^{(1)}\big(f(z) R/R_0\big)-\text{H}_0^{(1)}\big(f(-z) R/R_0\big)\Big]^2\\
        &-\eta^2\Big[f(z)\text{H}_1^{(1)}\big(f(z) R/R_0\big)\\
        &\qquad\qquad\quad-f(-z)\text{H}_1^{(1)}\big(f(-z) R/R_0\big)\Big]^2\Big\}\\
         & F_1(\boldsymbol{R}/R_0) =\text{Im} \int_0^\infty i \frac{dz}{z\sqrt{z^2+1}}\Big\{(z^2+1)\times\\
        & \Big[\text{H}_0^{(1)}\big(f(iz) R/R_0\big)-\text{H}_0^{(1)}\big(f(-iz) R/R_0\big)\Big]^2\\
        &+\eta^2\Big[f(iz)\text{H}_1^{(1)}\big(f(iz) R/R_0\big)\\
        &\qquad\qquad\quad-f(-iz)\text{H}_1^{(1)}\big(f(-iz) R/R_0\big)\Big]^2\Big\}
    \end{aligned}
\end{equation}
with $\eta=2(W/\Delta_0)(a/R_0)$, $R_0=\sqrt{2|m_1|/\Delta_0}$ and $f(z)=i\sqrt{k_0^2R_0^2-z}$. Other components of $\chi_{\alpha\beta}^{tt}(\boldsymbol{R})$ can be obtained in a similar way.

\section{Susceptibility at finite temperatures}\label{APP:VVFiniteT}

Our goal is to obtain the temperature dependence of the Bloembergen-Rowland exchange by computing the real-space static susceptibility \(\chi_{zz}(\boldsymbol R,T)=J_{zz}(\boldsymbol{R},T)/J_0\). We first treat thick TIs (no intersurface hybridization; gapless surface), and then consider thin TIs where hybridization opens a gap that sets the BR length scale.

\begin{figure}[t]
  \centering
  \includegraphics[width=0.5\linewidth]{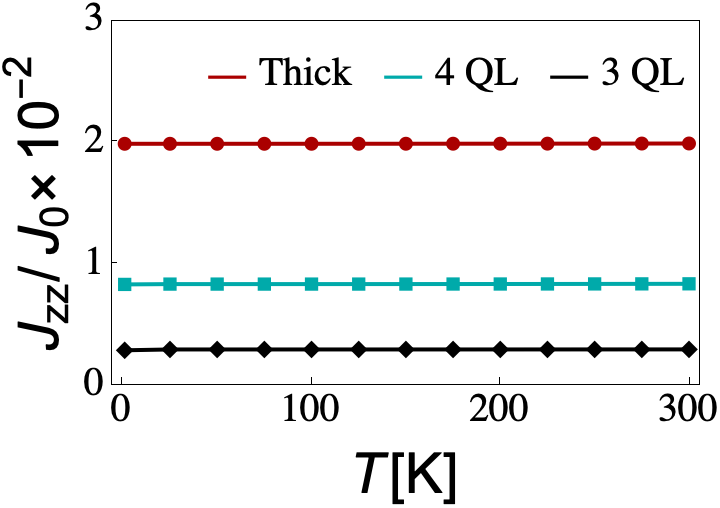}\hfill
  \caption{\justifying \(J_{zz}(R=a)\) versus \(T\) for a thick TI (\(\Delta=0\), gapless Dirac surface) (shown in red), 4 QL BST (shown in cyan) and 3 QL BST (shown in black). The temperature variation of the exchange coupling is less than a percent and 
  not visible on this plot.}
  \label{fig:ChiZT}
\end{figure}

For a thick TI the two surfaces do not hybridize. Since we are interested in temperatures near \(T_c\), we set the exchange gap to zero, \(\Delta=0\).
Starting from the imaginary-time susceptibility in Eq.~\eqref{Eq:SusceptibilityTau} and the momentum-space Matsubara Green's function
\[
G(\boldsymbol{k},i\omega_n)=\frac{i\omega_n\sigma_0+\hbar v_F(\boldsymbol{k}\times\boldsymbol{\sigma})_z}
{(i\omega_n)^2-\hbar^2 v_F^2 k^2},
\]
we write
\begin{equation}
\label{eq:GktauGapless}
G(\boldsymbol{k},\tau)=\frac{1}{\beta}\sum_{i\omega_n} 
\frac{e^{-i\omega_n\tau}\,\big[i\omega_n\sigma_0+\hbar v_F(\boldsymbol{k}\times\boldsymbol{\sigma})_z\big]}
{[i\omega_n-\epsilon_v(\boldsymbol{k})]\,[i\omega_n-\epsilon_c(\boldsymbol{k})]},
\end{equation}
with band energies \(\epsilon_{c(v)}(\boldsymbol{k})=\pm \hbar v_F |\boldsymbol{k}|\).
Evaluating the sum over poles of \(\tilde{n}_F(z)=[e^{-\beta z}+1]^{-1}\) gives
\begin{equation}
\label{Eq:GFktau}
\begin{aligned}
G(\boldsymbol{k},\tau) 
&=[\epsilon_v \tilde{n}_F(\epsilon_v)\,e^{-\epsilon_v\tau}-\epsilon_c \tilde{n}_F(\epsilon_c)\,e^{-\epsilon_c\tau}]\frac{\sigma_0}
{\epsilon_c-\epsilon_v}\, \\
& +[\tilde{n}_F(\epsilon_v)\,e^{-\epsilon_v\tau}-\tilde{n}_F(\epsilon_c)\,e^{-\epsilon_c\tau}]\frac{\hbar v_F(\boldsymbol{k}\times\boldsymbol{\sigma})_z}
{\epsilon_c-\epsilon_v} 
\end{aligned}
\end{equation}
Fourier transforming to real space (as in App.~\ref{App:RealSpaceGF}) yields
\begin{equation}
\label{eq:GRtauGapless}
\begin{aligned}
G(\boldsymbol{R},\tau)
&=\sum_{n=0}^{\infty}\Big[
\mathbb{F}^m_n(\boldsymbol{R})\,\sin(\omega_n\tau)\,\sigma_0 \\
&\qquad+\, i\,\mathbb{F}^m_n(\boldsymbol{R})\cos(\omega_n\tau)\,(\hat{\boldsymbol{R}}\times\boldsymbol{\sigma})_z
\Big]
\end{aligned}
\end{equation}
where \(\omega_n=(2n+1)\pi/\beta\) are fermionic Matsubara frequencies \cite{AGD1965Pergamon, Coleman2015} and
\[
\mathbb{F}^m_n(\boldsymbol{R}) \equiv -\frac{\omega_n}{\pi\beta W^2}\,
K_m\!\left(\frac{\omega_n}{W}\frac{R}{a}\right),\qquad m=0,1,
\]
with \(K_m\) the modified Bessel functions and \(W=\hbar v_F/a\).
Inserting Eq.~\eqref{eq:GRtauGapless} into Eq.~\eqref{Eq:SusceptibilityTau} and taking the spin trace gives
\begin{equation}
\label{eq:chiGaplessT}
\chi_{zz}(\boldsymbol{R},T)=\beta W\sum_{n=0}^{\infty}\Big([\mathbb{F}_n^0(\boldsymbol{R})]^2+[\mathbb{F}_n^1(\boldsymbol{R})]^2\Big),
\end{equation}
valid for all \(T\). Figure~\ref{fig:ChiZT} shows \(J_{zz}(R=a)\) versus temperature up to room temperature, in which the BR coupling is essentially unchanged (with a slight increase as conduction-band states become occupied). Extending the calculation to higher temperatures, up to \(T\!\sim\! W=\hbar v_F/a \approx 7000~\mathrm{K}\) \cite{BST2015,ou2023enhanced}, we find a shallow increase peaking near \(T\!\sim\!1000~\mathrm{K}\), followed by a gradual decrease. Overall, the variation in \(J_{zz}(T)\) remains negligible throughout the experimentally relevant regime \(T\!\ll\! W\).

For a thin TI, the two surfaces hybridize and the top-surface Green's function \(G_{tt}\) is (see Appendix \ref{App:BRIntTIThinFilms})

\begin{equation}\label{eq:GttkTau}
G_{tt}(\boldsymbol{k},\tau)=\frac{1}{\beta}\sum_{i\omega_n}
\frac{e^{-i\omega_n\tau}\,[i\omega_n\sigma_0+\hbar v_F(\boldsymbol{k}\times\boldsymbol{\sigma})_z]}
{[i\omega_n-\epsilon_v(\boldsymbol{k})]\,[i\omega_n-\epsilon_c(\boldsymbol{k})]},
\end{equation}
with band energies now gapped by hybridization \(\epsilon_{c(v)}(\boldsymbol{k})=\pm\sqrt{\hbar^2 v_F^2 k^2+(m_0-m_1 k^2)^2}\).
Proceeding as above and Fourier transforming, we obtain

\begin{equation}\label{eq:GRtauGapped}
\begin{aligned}
G_{tt}(\boldsymbol{R},\tau)
&=\sum_{n=0}^{\infty}\Big[
\mathbb{Y}_{n}(\boldsymbol{R})\,\sin(\omega_n\tau)\,\sigma_0 \\
&\qquad +\, i\,\mathbb{U}_{n}(\boldsymbol{R})\,\cos(\omega_n\tau)\,(\hat{\boldsymbol{R}}\times\boldsymbol{\sigma})_z
\Big],
\end{aligned}
\end{equation}
with
\begin{equation}
\label{eq:YnUnDefs}
\begin{aligned}
\mathbb{Y}_n(\boldsymbol{R}) &=\frac{(2\omega_n/\Delta_0)(a/R_0)^2}{\pi\beta\,\Delta_0\,\sqrt{1-(2\omega_n/\Delta_0)^2}}\times\\
&\!\left[
K_0\!\Big(\tfrac{R}{R_0}f_n^{+}\Big)
- K_0\!\Big(\tfrac{R}{R_0}f_n^{-}\Big)
\right],\\[4pt]
\mathbb{U}_n(\boldsymbol{R})&=\frac{2(W/\Delta_0)(a/R_0)^3}{\pi\beta\,\Delta_0\,\sqrt{1-(2\omega_n/\Delta_0)^2}}\times \\
&\!\left[
f_n^{+} K_1\!\Big(\tfrac{R}{R_0}f_n^{+}\Big)
- f_n^{-} K_1\!\Big(\tfrac{R}{R_0}f_n^{-}\Big)
\right],\\[4pt]
f_n^{\pm}&=\sqrt{k_0^2 R_0^2 \,\pm\, \sqrt{1-(2\omega_n/\Delta_0)^2}},
\end{aligned}
\end{equation}
where $\Delta_0=\frac{\hbar v_F}{|m_1|}\sqrt{\hbar^2 v_F^2-4 m_0 m_1}$, $R_0=\sqrt{\frac{2|m_1|}{\Delta_0}}$, and $k_0=\sqrt{(\hbar^2 v_F^2-2 m_0 m_1)/2 m_1^2}$ as discussed in Appendix \ref{App:BRIntTIThinFilms}. The top-surface susceptibility is then
\begin{equation}
\label{eq:chiGappedT}
\chi^{tt}_{zz}(\boldsymbol{R},T)=\beta W\sum_{n=0}^{\infty}
\Big[\mathbb{Y}_n^2(\boldsymbol{R})+\mathbb{U}_n^2(\boldsymbol{R})\Big].
\end{equation}
Figure~\ref{fig:ChiZT} shows \(J^{tt}_{zz}(R=a)\) as a function of temperature. 
Using parameters for 3--4\,QL BST, \(\Delta_0 \simeq 0.3~\mathrm{eV} \approx 3.5\times10^3~\mathrm{K}\) \cite{Wang2015}, the variation is \(<1\%\) up to room temperature. 
At higher temperatures, a shallow nonmonotonic trend appears: a slight increase up to \(T\!\approx\!1.6\times10^3~\mathrm{K}\) followed by a gradual decrease. 
Thus, for estimating the BR-mediated \(T_c\) enhancement, the \(T=0\) formalism is an excellent approximation.

\newpage

\bibliography{main}

\begin{thebibliography}{56}%
\makeatletter
\providecommand \@ifxundefined [1]{%
 \@ifx{#1\undefined}
}%
\providecommand \@ifnum [1]{%
 \ifnum #1\expandafter \@firstoftwo
 \else \expandafter \@secondoftwo
 \fi
}%
\providecommand \@ifx [1]{%
 \ifx #1\expandafter \@firstoftwo
 \else \expandafter \@secondoftwo
 \fi
}%
\providecommand \natexlab [1]{#1}%
\providecommand \enquote  [1]{``#1''}%
\providecommand \bibnamefont  [1]{#1}%
\providecommand \bibfnamefont [1]{#1}%
\providecommand \citenamefont [1]{#1}%
\providecommand \href@noop [0]{\@secondoftwo}%
\providecommand \href [0]{\begingroup \@sanitize@url \@href}%
\providecommand \@href[1]{\@@startlink{#1}\@@href}%
\providecommand \@@href[1]{\endgroup#1\@@endlink}%
\providecommand \@sanitize@url [0]{\catcode `\\12\catcode `\$12\catcode `\&12\catcode `\#12\catcode `\^12\catcode `\_12\catcode `\%12\relax}%
\providecommand \@@startlink[1]{}%
\providecommand \@@endlink[0]{}%
\providecommand \url  [0]{\begingroup\@sanitize@url \@url }%
\providecommand \@url [1]{\endgroup\@href {#1}{\urlprefix }}%
\providecommand \urlprefix  [0]{URL }%
\providecommand \Eprint [0]{\href }%
\providecommand \doibase [0]{https://doi.org/}%
\providecommand \selectlanguage [0]{\@gobble}%
\providecommand \bibinfo  [0]{\@secondoftwo}%
\providecommand \bibfield  [0]{\@secondoftwo}%
\providecommand \translation [1]{[#1]}%
\providecommand \BibitemOpen [0]{}%
\providecommand \bibitemStop [0]{}%
\providecommand \bibitemNoStop [0]{.\EOS\space}%
\providecommand \EOS [0]{\spacefactor3000\relax}%
\providecommand \BibitemShut  [1]{\csname bibitem#1\endcsname}%
\let\auto@bib@innerbib\@empty
\bibitem [{\citenamefont {Zhang}\ \emph {et~al.}(2013)\citenamefont {Zhang}, \citenamefont {Chang}, \citenamefont {Tang}, \citenamefont {Zhang}, \citenamefont {Feng}, \citenamefont {Li}, \citenamefont {li~Wang}, \citenamefont {Chen}, \citenamefont {Liu}, \citenamefont {Duan}, \citenamefont {He}, \citenamefont {Xue}, \citenamefont {Ma},\ and\ \citenamefont {Wang}}]{Zhang2013}%
  \BibitemOpen
  \bibfield  {author} {\bibinfo {author} {\bibfnamefont {J.}~\bibnamefont {Zhang}}, \bibinfo {author} {\bibfnamefont {C.-Z.}\ \bibnamefont {Chang}}, \bibinfo {author} {\bibfnamefont {P.}~\bibnamefont {Tang}}, \bibinfo {author} {\bibfnamefont {Z.}~\bibnamefont {Zhang}}, \bibinfo {author} {\bibfnamefont {X.}~\bibnamefont {Feng}}, \bibinfo {author} {\bibfnamefont {K.}~\bibnamefont {Li}}, \bibinfo {author} {\bibfnamefont {L.}~\bibnamefont {li~Wang}}, \bibinfo {author} {\bibfnamefont {X.}~\bibnamefont {Chen}}, \bibinfo {author} {\bibfnamefont {C.}~\bibnamefont {Liu}}, \bibinfo {author} {\bibfnamefont {W.}~\bibnamefont {Duan}}, \bibinfo {author} {\bibfnamefont {K.}~\bibnamefont {He}}, \bibinfo {author} {\bibfnamefont {Q.-K.}\ \bibnamefont {Xue}}, \bibinfo {author} {\bibfnamefont {X.}~\bibnamefont {Ma}},\ and\ \bibinfo {author} {\bibfnamefont {Y.}~\bibnamefont {Wang}},\ }\bibfield  {title} {\bibinfo {title} {Topology-driven magnetic quantum phase transition in topological insulators},\ }\href
  {https://doi.org/10.1126/science.1230905} {\bibfield  {journal} {\bibinfo  {journal} {Science}\ }\textbf {\bibinfo {volume} {339}},\ \bibinfo {pages} {1582} (\bibinfo {year} {2013})}\BibitemShut {NoStop}%
\bibitem [{\citenamefont {Tokura}\ \emph {et~al.}(2019)\citenamefont {Tokura}, \citenamefont {Yasuda},\ and\ \citenamefont {Tsukazaki}}]{tokura2019}%
  \BibitemOpen
  \bibfield  {author} {\bibinfo {author} {\bibfnamefont {Y.}~\bibnamefont {Tokura}}, \bibinfo {author} {\bibfnamefont {K.}~\bibnamefont {Yasuda}},\ and\ \bibinfo {author} {\bibfnamefont {A.}~\bibnamefont {Tsukazaki}},\ }\bibfield  {title} {\bibinfo {title} {Magnetic topological insulators},\ }\href {https://doi.org/10.1038/s42254-018-0011-5} {\bibfield  {journal} {\bibinfo  {journal} {Nature Reviews Physics}\ }\textbf {\bibinfo {volume} {1}},\ \bibinfo {pages} {126} (\bibinfo {year} {2019})}\BibitemShut {NoStop}%
\bibitem [{\citenamefont {Bernevig}\ \emph {et~al.}(2022)\citenamefont {Bernevig}, \citenamefont {Felser},\ and\ \citenamefont {Beidenkopf}}]{Bernevig_2022}%
  \BibitemOpen
  \bibfield  {author} {\bibinfo {author} {\bibfnamefont {B.~A.}\ \bibnamefont {Bernevig}}, \bibinfo {author} {\bibfnamefont {C.}~\bibnamefont {Felser}},\ and\ \bibinfo {author} {\bibfnamefont {H.}~\bibnamefont {Beidenkopf}},\ }\bibfield  {title} {\bibinfo {title} {Progress and prospects in magnetic topological materials},\ }\href {https://doi.org/10.1038/s41586-021-04105-x} {\bibfield  {journal} {\bibinfo  {journal} {Nature}\ }\textbf {\bibinfo {volume} {603}},\ \bibinfo {pages} {41–51} (\bibinfo {year} {2022})}\BibitemShut {NoStop}%
\bibitem [{\citenamefont {Yu}\ \emph {et~al.}(2010)\citenamefont {Yu}, \citenamefont {Zhang}, \citenamefont {Zhang}, \citenamefont {Zhang}, \citenamefont {Dai},\ and\ \citenamefont {Fang}}]{yu2010QAHE}%
  \BibitemOpen
  \bibfield  {author} {\bibinfo {author} {\bibfnamefont {R.}~\bibnamefont {Yu}}, \bibinfo {author} {\bibfnamefont {W.}~\bibnamefont {Zhang}}, \bibinfo {author} {\bibfnamefont {H.-J.}\ \bibnamefont {Zhang}}, \bibinfo {author} {\bibfnamefont {S.-C.}\ \bibnamefont {Zhang}}, \bibinfo {author} {\bibfnamefont {X.}~\bibnamefont {Dai}},\ and\ \bibinfo {author} {\bibfnamefont {Z.}~\bibnamefont {Fang}},\ }\bibfield  {title} {\bibinfo {title} {Quantized anomalous hall effect in magnetic topological insulators},\ }\href {https://doi.org/10.1126/science.1187485} {\bibfield  {journal} {\bibinfo  {journal} {Science}\ }\textbf {\bibinfo {volume} {329}},\ \bibinfo {pages} {61} (\bibinfo {year} {2010})}\BibitemShut {NoStop}%
\bibitem [{\citenamefont {Chang}\ \emph {et~al.}(2013)\citenamefont {Chang}, \citenamefont {Zhang}, \citenamefont {Feng}, \citenamefont {Shen}, \citenamefont {Zhang}, \citenamefont {Guo}, \citenamefont {Li}, \citenamefont {Ou}, \citenamefont {Wei}, \citenamefont {Wang}, \citenamefont {Ji}, \citenamefont {Feng}, \citenamefont {Ji}, \citenamefont {Chen}, \citenamefont {Jia}, \citenamefont {Dai}, \citenamefont {Fang}, \citenamefont {Zhang}, \citenamefont {He}, \citenamefont {Wang}, \citenamefont {Lu}, \citenamefont {Ma},\ and\ \citenamefont {Xue}}]{ChangQAHE2013}%
  \BibitemOpen
  \bibfield  {author} {\bibinfo {author} {\bibfnamefont {C.-Z.}\ \bibnamefont {Chang}}, \bibinfo {author} {\bibfnamefont {J.}~\bibnamefont {Zhang}}, \bibinfo {author} {\bibfnamefont {X.}~\bibnamefont {Feng}}, \bibinfo {author} {\bibfnamefont {J.}~\bibnamefont {Shen}}, \bibinfo {author} {\bibfnamefont {Z.}~\bibnamefont {Zhang}}, \bibinfo {author} {\bibfnamefont {M.}~\bibnamefont {Guo}}, \bibinfo {author} {\bibfnamefont {K.}~\bibnamefont {Li}}, \bibinfo {author} {\bibfnamefont {Y.}~\bibnamefont {Ou}}, \bibinfo {author} {\bibfnamefont {P.}~\bibnamefont {Wei}}, \bibinfo {author} {\bibfnamefont {L.-L.}\ \bibnamefont {Wang}}, \bibinfo {author} {\bibfnamefont {Z.-Q.}\ \bibnamefont {Ji}}, \bibinfo {author} {\bibfnamefont {Y.}~\bibnamefont {Feng}}, \bibinfo {author} {\bibfnamefont {S.}~\bibnamefont {Ji}}, \bibinfo {author} {\bibfnamefont {X.}~\bibnamefont {Chen}}, \bibinfo {author} {\bibfnamefont {J.}~\bibnamefont {Jia}}, \bibinfo {author} {\bibfnamefont {X.}~\bibnamefont {Dai}}, \bibinfo {author} {\bibfnamefont
  {Z.}~\bibnamefont {Fang}}, \bibinfo {author} {\bibfnamefont {S.-C.}\ \bibnamefont {Zhang}}, \bibinfo {author} {\bibfnamefont {K.}~\bibnamefont {He}}, \bibinfo {author} {\bibfnamefont {Y.}~\bibnamefont {Wang}}, \bibinfo {author} {\bibfnamefont {L.}~\bibnamefont {Lu}}, \bibinfo {author} {\bibfnamefont {X.-C.}\ \bibnamefont {Ma}},\ and\ \bibinfo {author} {\bibfnamefont {Q.-K.}\ \bibnamefont {Xue}},\ }\bibfield  {title} {\bibinfo {title} {Experimental observation of the quantum anomalous hall effect in a magnetic topological insulator},\ }\href {https://doi.org/10.1126/science.1234414} {\bibfield  {journal} {\bibinfo  {journal} {Science}\ }\textbf {\bibinfo {volume} {340}},\ \bibinfo {pages} {167} (\bibinfo {year} {2013})}\BibitemShut {NoStop}%
\bibitem [{\citenamefont {Chang}\ \emph {et~al.}(2015)\citenamefont {Chang}, \citenamefont {Zhao}, \citenamefont {Kim}, \citenamefont {Zhang}, \citenamefont {Assaf}, \citenamefont {Heiman}, \citenamefont {Zhang}, \citenamefont {Liu}, \citenamefont {Chan},\ and\ \citenamefont {Moodera}}]{ChangQAHE2015}%
  \BibitemOpen
  \bibfield  {author} {\bibinfo {author} {\bibfnamefont {C.-Z.}\ \bibnamefont {Chang}}, \bibinfo {author} {\bibfnamefont {W.}~\bibnamefont {Zhao}}, \bibinfo {author} {\bibfnamefont {D.~Y.}\ \bibnamefont {Kim}}, \bibinfo {author} {\bibfnamefont {H.}~\bibnamefont {Zhang}}, \bibinfo {author} {\bibfnamefont {B.~A.}\ \bibnamefont {Assaf}}, \bibinfo {author} {\bibfnamefont {D.}~\bibnamefont {Heiman}}, \bibinfo {author} {\bibfnamefont {S.-C.}\ \bibnamefont {Zhang}}, \bibinfo {author} {\bibfnamefont {C.}~\bibnamefont {Liu}}, \bibinfo {author} {\bibfnamefont {M.~H.~W.}\ \bibnamefont {Chan}},\ and\ \bibinfo {author} {\bibfnamefont {J.~S.}\ \bibnamefont {Moodera}},\ }\bibfield  {title} {\bibinfo {title} {High-precision realization of robust quantum anomalous hall state in a hard ferromagnetic topological insulator},\ }\href {https://doi.org/10.1038/nmat4204} {\bibfield  {journal} {\bibinfo  {journal} {Nature Materials}\ }\textbf {\bibinfo {volume} {14}},\ \bibinfo {pages} {473} (\bibinfo {year} {2015})}\BibitemShut
  {NoStop}%
\bibitem [{\citenamefont {Vobornik}\ \emph {et~al.}(2011)\citenamefont {Vobornik}, \citenamefont {Manju}, \citenamefont {Fujii}, \citenamefont {Borgatti}, \citenamefont {Torelli}, \citenamefont {Krizmancic}, \citenamefont {Hor}, \citenamefont {Cava},\ and\ \citenamefont {Panaccione}}]{Vobornik2011}%
  \BibitemOpen
  \bibfield  {author} {\bibinfo {author} {\bibfnamefont {I.}~\bibnamefont {Vobornik}}, \bibinfo {author} {\bibfnamefont {U.}~\bibnamefont {Manju}}, \bibinfo {author} {\bibfnamefont {J.}~\bibnamefont {Fujii}}, \bibinfo {author} {\bibfnamefont {F.}~\bibnamefont {Borgatti}}, \bibinfo {author} {\bibfnamefont {P.}~\bibnamefont {Torelli}}, \bibinfo {author} {\bibfnamefont {D.}~\bibnamefont {Krizmancic}}, \bibinfo {author} {\bibfnamefont {Y.~S.}\ \bibnamefont {Hor}}, \bibinfo {author} {\bibfnamefont {R.~J.}\ \bibnamefont {Cava}},\ and\ \bibinfo {author} {\bibfnamefont {G.}~\bibnamefont {Panaccione}},\ }\bibfield  {title} {\bibinfo {title} {Magnetic proximity effect as a pathway to spintronic applications of topological insulators},\ }\href {https://doi.org/10.1021/nl201275q} {\bibfield  {journal} {\bibinfo  {journal} {Nano Letters}\ }\textbf {\bibinfo {volume} {11}},\ \bibinfo {pages} {4079} (\bibinfo {year} {2011})}\BibitemShut {NoStop}%
\bibitem [{\citenamefont {Wei}\ \emph {et~al.}(2013)\citenamefont {Wei}, \citenamefont {Katmis}, \citenamefont {Assaf}, \citenamefont {Steinberg}, \citenamefont {Jarillo-Herrero}, \citenamefont {Heiman},\ and\ \citenamefont {Moodera}}]{Wei2013}%
  \BibitemOpen
  \bibfield  {author} {\bibinfo {author} {\bibfnamefont {P.}~\bibnamefont {Wei}}, \bibinfo {author} {\bibfnamefont {F.}~\bibnamefont {Katmis}}, \bibinfo {author} {\bibfnamefont {B.~A.}\ \bibnamefont {Assaf}}, \bibinfo {author} {\bibfnamefont {H.}~\bibnamefont {Steinberg}}, \bibinfo {author} {\bibfnamefont {P.}~\bibnamefont {Jarillo-Herrero}}, \bibinfo {author} {\bibfnamefont {D.}~\bibnamefont {Heiman}},\ and\ \bibinfo {author} {\bibfnamefont {J.~S.}\ \bibnamefont {Moodera}},\ }\bibfield  {title} {\bibinfo {title} {Exchange-coupling-induced symmetry breaking in topological insulators},\ }\href {https://doi.org/10.1103/PhysRevLett.110.186807} {\bibfield  {journal} {\bibinfo  {journal} {Phys. Rev. Lett.}\ }\textbf {\bibinfo {volume} {110}},\ \bibinfo {pages} {186807} (\bibinfo {year} {2013})}\BibitemShut {NoStop}%
\bibitem [{\citenamefont {Lang}\ \emph {et~al.}(2014)\citenamefont {Lang}, \citenamefont {Montazeri}, \citenamefont {Onbasli}, \citenamefont {Kou}, \citenamefont {Fan}, \citenamefont {Upadhyaya}, \citenamefont {Yao}, \citenamefont {Liu}, \citenamefont {Jiang}, \citenamefont {Jiang}, \citenamefont {Wong}, \citenamefont {Yu}, \citenamefont {Tang}, \citenamefont {Nie}, \citenamefont {He}, \citenamefont {Schwartz}, \citenamefont {Wang}, \citenamefont {Ross},\ and\ \citenamefont {Wang}}]{Lang2014}%
  \BibitemOpen
  \bibfield  {author} {\bibinfo {author} {\bibfnamefont {M.}~\bibnamefont {Lang}}, \bibinfo {author} {\bibfnamefont {M.}~\bibnamefont {Montazeri}}, \bibinfo {author} {\bibfnamefont {M.~C.}\ \bibnamefont {Onbasli}}, \bibinfo {author} {\bibfnamefont {X.}~\bibnamefont {Kou}}, \bibinfo {author} {\bibfnamefont {Y.}~\bibnamefont {Fan}}, \bibinfo {author} {\bibfnamefont {P.}~\bibnamefont {Upadhyaya}}, \bibinfo {author} {\bibfnamefont {K.}~\bibnamefont {Yao}}, \bibinfo {author} {\bibfnamefont {F.}~\bibnamefont {Liu}}, \bibinfo {author} {\bibfnamefont {Y.}~\bibnamefont {Jiang}}, \bibinfo {author} {\bibfnamefont {W.}~\bibnamefont {Jiang}}, \bibinfo {author} {\bibfnamefont {K.~L.}\ \bibnamefont {Wong}}, \bibinfo {author} {\bibfnamefont {G.}~\bibnamefont {Yu}}, \bibinfo {author} {\bibfnamefont {J.}~\bibnamefont {Tang}}, \bibinfo {author} {\bibfnamefont {T.}~\bibnamefont {Nie}}, \bibinfo {author} {\bibfnamefont {L.}~\bibnamefont {He}}, \bibinfo {author} {\bibfnamefont {R.~N.}\ \bibnamefont {Schwartz}}, \bibinfo {author}
  {\bibfnamefont {Y.}~\bibnamefont {Wang}}, \bibinfo {author} {\bibfnamefont {C.~A.}\ \bibnamefont {Ross}},\ and\ \bibinfo {author} {\bibfnamefont {K.~L.}\ \bibnamefont {Wang}},\ }\bibfield  {title} {\bibinfo {title} {Proximity induced high-temperature magnetic order in topological insulator - ferrimagnetic insulator heterostructure},\ }\href {https://doi.org/10.1021/nl500973k} {\bibfield  {journal} {\bibinfo  {journal} {Nano Letters}\ }\textbf {\bibinfo {volume} {14}},\ \bibinfo {pages} {3459} (\bibinfo {year} {2014})}\BibitemShut {NoStop}%
\bibitem [{\citenamefont {Jiang}\ \emph {et~al.}(2015)\citenamefont {Jiang}, \citenamefont {Chang}, \citenamefont {Tang}, \citenamefont {Wei}, \citenamefont {Moodera},\ and\ \citenamefont {Shi}}]{Jiang2015}%
  \BibitemOpen
  \bibfield  {author} {\bibinfo {author} {\bibfnamefont {Z.}~\bibnamefont {Jiang}}, \bibinfo {author} {\bibfnamefont {C.-Z.}\ \bibnamefont {Chang}}, \bibinfo {author} {\bibfnamefont {C.}~\bibnamefont {Tang}}, \bibinfo {author} {\bibfnamefont {P.}~\bibnamefont {Wei}}, \bibinfo {author} {\bibfnamefont {J.~S.}\ \bibnamefont {Moodera}},\ and\ \bibinfo {author} {\bibfnamefont {J.}~\bibnamefont {Shi}},\ }\bibfield  {title} {\bibinfo {title} {Independent tuning of electronic properties and induced ferromagnetism in topological insulators with heterostructure approach},\ }\href {https://doi.org/10.1021/acs.nanolett.5b01905} {\bibfield  {journal} {\bibinfo  {journal} {Nano Letters}\ }\textbf {\bibinfo {volume} {15}},\ \bibinfo {pages} {5835} (\bibinfo {year} {2015})}\BibitemShut {NoStop}%
\bibitem [{\citenamefont {Tse}\ and\ \citenamefont {MacDonald}(2010)}]{TseMOKEandFaraday}%
  \BibitemOpen
  \bibfield  {author} {\bibinfo {author} {\bibfnamefont {W.-K.}\ \bibnamefont {Tse}}\ and\ \bibinfo {author} {\bibfnamefont {A.~H.}\ \bibnamefont {MacDonald}},\ }\bibfield  {title} {\bibinfo {title} {Giant magneto-optical kerr effect and universal faraday effect in thin-film topological insulators},\ }\href {https://doi.org/10.1103/PhysRevLett.105.057401} {\bibfield  {journal} {\bibinfo  {journal} {Phys. Rev. Lett.}\ }\textbf {\bibinfo {volume} {105}},\ \bibinfo {pages} {057401} (\bibinfo {year} {2010})}\BibitemShut {NoStop}%
\bibitem [{\citenamefont {Mogi}\ \emph {et~al.}(2022)\citenamefont {Mogi}, \citenamefont {Okamura}, \citenamefont {Kawamura}, \citenamefont {Yoshimi}, \citenamefont {Yasuda}, \citenamefont {Tsukazaki}, \citenamefont {Takahashi}, \citenamefont {Morimoto}, \citenamefont {Nagaosa}, \citenamefont {Kawasaki}, \citenamefont {Takahashi},\ and\ \citenamefont {Tokura}}]{Mogi_2022}%
  \BibitemOpen
  \bibfield  {author} {\bibinfo {author} {\bibfnamefont {M.}~\bibnamefont {Mogi}}, \bibinfo {author} {\bibfnamefont {Y.}~\bibnamefont {Okamura}}, \bibinfo {author} {\bibfnamefont {M.}~\bibnamefont {Kawamura}}, \bibinfo {author} {\bibfnamefont {R.}~\bibnamefont {Yoshimi}}, \bibinfo {author} {\bibfnamefont {K.}~\bibnamefont {Yasuda}}, \bibinfo {author} {\bibfnamefont {A.}~\bibnamefont {Tsukazaki}}, \bibinfo {author} {\bibfnamefont {K.~S.}\ \bibnamefont {Takahashi}}, \bibinfo {author} {\bibfnamefont {T.}~\bibnamefont {Morimoto}}, \bibinfo {author} {\bibfnamefont {N.}~\bibnamefont {Nagaosa}}, \bibinfo {author} {\bibfnamefont {M.}~\bibnamefont {Kawasaki}}, \bibinfo {author} {\bibfnamefont {Y.}~\bibnamefont {Takahashi}},\ and\ \bibinfo {author} {\bibfnamefont {Y.}~\bibnamefont {Tokura}},\ }\bibfield  {title} {\bibinfo {title} {Experimental signature of the parity anomaly in a semi-magnetic topological insulator},\ }\href {https://doi.org/10.1038/s41567-021-01490-y} {\bibfield  {journal} {\bibinfo  {journal} {Nature
  Physics}\ }\textbf {\bibinfo {volume} {18}},\ \bibinfo {pages} {390–394} (\bibinfo {year} {2022})}\BibitemShut {NoStop}%
\bibitem [{\citenamefont {Jain}\ \emph {et~al.}(2024)\citenamefont {Jain}, \citenamefont {Roddy}, \citenamefont {Gupta}, \citenamefont {Huang}, \citenamefont {Sayeed}, \citenamefont {Alnaser}, \citenamefont {Vashist}, \citenamefont {Watanabe}, \citenamefont {Taniguchi}, \citenamefont {Deshpande}, \citenamefont {Sparks},\ and\ \citenamefont {Ralph}}]{Jain2024QAHE}%
  \BibitemOpen
  \bibfield  {author} {\bibinfo {author} {\bibfnamefont {R.}~\bibnamefont {Jain}}, \bibinfo {author} {\bibfnamefont {M.}~\bibnamefont {Roddy}}, \bibinfo {author} {\bibfnamefont {V.}~\bibnamefont {Gupta}}, \bibinfo {author} {\bibfnamefont {B.}~\bibnamefont {Huang}}, \bibinfo {author} {\bibfnamefont {H.~M.}\ \bibnamefont {Sayeed}}, \bibinfo {author} {\bibfnamefont {H.~F.}\ \bibnamefont {Alnaser}}, \bibinfo {author} {\bibfnamefont {A.}~\bibnamefont {Vashist}}, \bibinfo {author} {\bibfnamefont {K.}~\bibnamefont {Watanabe}}, \bibinfo {author} {\bibfnamefont {T.}~\bibnamefont {Taniguchi}}, \bibinfo {author} {\bibfnamefont {V.~V.}\ \bibnamefont {Deshpande}}, \bibinfo {author} {\bibfnamefont {T.~D.}\ \bibnamefont {Sparks}},\ and\ \bibinfo {author} {\bibfnamefont {D.~C.}\ \bibnamefont {Ralph}},\ }\href@noop {} {\bibinfo {title} {A quantized anomalous hall effect above 4.2 k in stacked topological insulator/magnet bilayers}} (\bibinfo {year} {2024}),\ \Eprint {https://arxiv.org/abs/2412.05380} {arXiv:2412.05380
  [cond-mat.mtrl-sci]} \BibitemShut {NoStop}%
\bibitem [{\citenamefont {Chi}\ and\ \citenamefont {Moodera}(2022)}]{Chi2022}%
  \BibitemOpen
  \bibfield  {author} {\bibinfo {author} {\bibfnamefont {H.}~\bibnamefont {Chi}}\ and\ \bibinfo {author} {\bibfnamefont {J.~S.}\ \bibnamefont {Moodera}},\ }\bibfield  {title} {\bibinfo {title} {{Progress and prospects in the quantum anomalous Hall effect}},\ }\href {https://doi.org/10.1063/5.0100989} {\bibfield  {journal} {\bibinfo  {journal} {APL Materials}\ }\textbf {\bibinfo {volume} {10}},\ \bibinfo {pages} {090903} (\bibinfo {year} {2022})}\BibitemShut {NoStop}%
\bibitem [{\citenamefont {Chang}\ \emph {et~al.}(2023)\citenamefont {Chang}, \citenamefont {Liu},\ and\ \citenamefont {MacDonald}}]{chang2023colloquium}%
  \BibitemOpen
  \bibfield  {author} {\bibinfo {author} {\bibfnamefont {C.-Z.}\ \bibnamefont {Chang}}, \bibinfo {author} {\bibfnamefont {C.-X.}\ \bibnamefont {Liu}},\ and\ \bibinfo {author} {\bibfnamefont {A.~H.}\ \bibnamefont {MacDonald}},\ }\bibfield  {title} {\bibinfo {title} {Colloquium: Quantum anomalous hall effect},\ }\href {https://doi.org/10.1103/RevModPhys.95.011002} {\bibfield  {journal} {\bibinfo  {journal} {Rev. Mod. Phys.}\ }\textbf {\bibinfo {volume} {95}},\ \bibinfo {pages} {011002} (\bibinfo {year} {2023})}\BibitemShut {NoStop}%
\bibitem [{\citenamefont {Ou}\ \emph {et~al.}(2025)\citenamefont {Ou}, \citenamefont {Mirzhalilov}, \citenamefont {Nemes}, \citenamefont {Martinez}, \citenamefont {Rocci}, \citenamefont {Duong}, \citenamefont {Akey}, \citenamefont {Foucher}, \citenamefont {Ge}, \citenamefont {Suri}, \citenamefont {Wang}, \citenamefont {Ambaye}, \citenamefont {Keum}, \citenamefont {Randeria}, \citenamefont {Trivedi}, \citenamefont {Burch}, \citenamefont {Bell}, \citenamefont {Ross}, \citenamefont {Wu}, \citenamefont {Heiman}, \citenamefont {Lauter}, \citenamefont {Moodera},\ and\ \citenamefont {Chi}}]{ou2023enhanced}%
  \BibitemOpen
  \bibfield  {author} {\bibinfo {author} {\bibfnamefont {Y.}~\bibnamefont {Ou}}, \bibinfo {author} {\bibfnamefont {M.}~\bibnamefont {Mirzhalilov}}, \bibinfo {author} {\bibfnamefont {N.~M.}\ \bibnamefont {Nemes}}, \bibinfo {author} {\bibfnamefont {J.~L.}\ \bibnamefont {Martinez}}, \bibinfo {author} {\bibfnamefont {M.}~\bibnamefont {Rocci}}, \bibinfo {author} {\bibfnamefont {A.}~\bibnamefont {Duong}}, \bibinfo {author} {\bibfnamefont {A.}~\bibnamefont {Akey}}, \bibinfo {author} {\bibfnamefont {A.~C.}\ \bibnamefont {Foucher}}, \bibinfo {author} {\bibfnamefont {W.}~\bibnamefont {Ge}}, \bibinfo {author} {\bibfnamefont {D.}~\bibnamefont {Suri}}, \bibinfo {author} {\bibfnamefont {Y.}~\bibnamefont {Wang}}, \bibinfo {author} {\bibfnamefont {H.}~\bibnamefont {Ambaye}}, \bibinfo {author} {\bibfnamefont {J.}~\bibnamefont {Keum}}, \bibinfo {author} {\bibfnamefont {M.}~\bibnamefont {Randeria}}, \bibinfo {author} {\bibfnamefont {N.}~\bibnamefont {Trivedi}}, \bibinfo {author} {\bibfnamefont {K.~S.}\ \bibnamefont {Burch}},
  \bibinfo {author} {\bibfnamefont {D.~C.}\ \bibnamefont {Bell}}, \bibinfo {author} {\bibfnamefont {F.~M.}\ \bibnamefont {Ross}}, \bibinfo {author} {\bibfnamefont {W.}~\bibnamefont {Wu}}, \bibinfo {author} {\bibfnamefont {D.}~\bibnamefont {Heiman}}, \bibinfo {author} {\bibfnamefont {V.}~\bibnamefont {Lauter}}, \bibinfo {author} {\bibfnamefont {J.~S.}\ \bibnamefont {Moodera}},\ and\ \bibinfo {author} {\bibfnamefont {H.}~\bibnamefont {Chi}},\ }\bibfield  {title} {\bibinfo {title} {Enhanced ferromagnetism in monolayer {Cr$_2$Te$_3$} via topological insulator coupling},\ }\href {https://doi.org/10.1088/1361-6633/add9c5} {\bibfield  {journal} {\bibinfo  {journal} {Reports on Progress in Physics}\ }\textbf {\bibinfo {volume} {88}},\ \bibinfo {pages} {060501} (\bibinfo {year} {2025})}\BibitemShut {NoStop}%
\bibitem [{\citenamefont {Alegria}\ \emph {et~al.}(2014)\citenamefont {Alegria}, \citenamefont {Ji}, \citenamefont {Yao}, \citenamefont {Clarke}, \citenamefont {Cava},\ and\ \citenamefont {Petta}}]{CGT2014}%
  \BibitemOpen
  \bibfield  {author} {\bibinfo {author} {\bibfnamefont {L.~D.}\ \bibnamefont {Alegria}}, \bibinfo {author} {\bibfnamefont {H.}~\bibnamefont {Ji}}, \bibinfo {author} {\bibfnamefont {N.}~\bibnamefont {Yao}}, \bibinfo {author} {\bibfnamefont {J.~J.}\ \bibnamefont {Clarke}}, \bibinfo {author} {\bibfnamefont {R.~J.}\ \bibnamefont {Cava}},\ and\ \bibinfo {author} {\bibfnamefont {J.~R.}\ \bibnamefont {Petta}},\ }\bibfield  {title} {\bibinfo {title} {Large anomalous hall effect in ferromagnetic insulator-topological insulator heterostructures},\ }\href {https://doi.org/10.1063/1.4892353} {\bibfield  {journal} {\bibinfo  {journal} {Applied Physics Letters}\ }\textbf {\bibinfo {volume} {105}},\ \bibinfo {pages} {053512} (\bibinfo {year} {2014})}\BibitemShut {NoStop}%
\bibitem [{\citenamefont {Katmis}\ \emph {et~al.}(2016)\citenamefont {Katmis}, \citenamefont {Lauter}, \citenamefont {Nogueira}, \citenamefont {Assaf}, \citenamefont {Jamer}, \citenamefont {Wei}, \citenamefont {Satpati}, \citenamefont {Freeland}, \citenamefont {Eremin}, \citenamefont {Heiman}, \citenamefont {Jarillo-Herrero},\ and\ \citenamefont {Moodera}}]{EuS2016}%
  \BibitemOpen
  \bibfield  {author} {\bibinfo {author} {\bibfnamefont {F.}~\bibnamefont {Katmis}}, \bibinfo {author} {\bibfnamefont {V.}~\bibnamefont {Lauter}}, \bibinfo {author} {\bibfnamefont {F.~S.}\ \bibnamefont {Nogueira}}, \bibinfo {author} {\bibfnamefont {B.~A.}\ \bibnamefont {Assaf}}, \bibinfo {author} {\bibfnamefont {M.~E.}\ \bibnamefont {Jamer}}, \bibinfo {author} {\bibfnamefont {P.}~\bibnamefont {Wei}}, \bibinfo {author} {\bibfnamefont {B.}~\bibnamefont {Satpati}}, \bibinfo {author} {\bibfnamefont {J.~W.}\ \bibnamefont {Freeland}}, \bibinfo {author} {\bibfnamefont {I.}~\bibnamefont {Eremin}}, \bibinfo {author} {\bibfnamefont {D.}~\bibnamefont {Heiman}}, \bibinfo {author} {\bibfnamefont {P.}~\bibnamefont {Jarillo-Herrero}},\ and\ \bibinfo {author} {\bibfnamefont {J.~S.}\ \bibnamefont {Moodera}},\ }\bibfield  {title} {\bibinfo {title} {{A high-temperature ferromagnetic topological insulating phase by proximity coupling}},\ }\href {https://doi.org/10.1038/nature17635} {\bibfield  {journal} {\bibinfo  {journal}
  {Nature}\ }\textbf {\bibinfo {volume} {533}},\ \bibinfo {pages} {513} (\bibinfo {year} {2016})}\BibitemShut {NoStop}%
\bibitem [{\citenamefont {Liu}\ \emph {et~al.}(2020)\citenamefont {Liu}, \citenamefont {Kally}, \citenamefont {Pillsbury}, \citenamefont {Liu}, \citenamefont {Chang}, \citenamefont {Ding}, \citenamefont {Cheng}, \citenamefont {Hilse}, \citenamefont {Engel-Herbert}, \citenamefont {Richardella}, \citenamefont {Samarth},\ and\ \citenamefont {Wu}}]{YIG/TI}%
  \BibitemOpen
  \bibfield  {author} {\bibinfo {author} {\bibfnamefont {T.}~\bibnamefont {Liu}}, \bibinfo {author} {\bibfnamefont {J.}~\bibnamefont {Kally}}, \bibinfo {author} {\bibfnamefont {T.}~\bibnamefont {Pillsbury}}, \bibinfo {author} {\bibfnamefont {C.}~\bibnamefont {Liu}}, \bibinfo {author} {\bibfnamefont {H.}~\bibnamefont {Chang}}, \bibinfo {author} {\bibfnamefont {J.}~\bibnamefont {Ding}}, \bibinfo {author} {\bibfnamefont {Y.}~\bibnamefont {Cheng}}, \bibinfo {author} {\bibfnamefont {M.}~\bibnamefont {Hilse}}, \bibinfo {author} {\bibfnamefont {R.}~\bibnamefont {Engel-Herbert}}, \bibinfo {author} {\bibfnamefont {A.}~\bibnamefont {Richardella}}, \bibinfo {author} {\bibfnamefont {N.}~\bibnamefont {Samarth}},\ and\ \bibinfo {author} {\bibfnamefont {M.}~\bibnamefont {Wu}},\ }\bibfield  {title} {\bibinfo {title} {Changes of magnetism in a magnetic insulator due to proximity to a topological insulator},\ }\href {https://doi.org/10.1103/PhysRevLett.125.017204} {\bibfield  {journal} {\bibinfo  {journal} {Phys. Rev. Lett.}\
  }\textbf {\bibinfo {volume} {125}},\ \bibinfo {pages} {017204} (\bibinfo {year} {2020})}\BibitemShut {NoStop}%
\bibitem [{\citenamefont {Zhu}\ \emph {et~al.}(2018)\citenamefont {Zhu}, \citenamefont {Meng}, \citenamefont {Liang}, \citenamefont {Shi}, \citenamefont {Zhao}, \citenamefont {Cheng}, \citenamefont {Li}, \citenamefont {Zhai}, \citenamefont {Lu}, \citenamefont {Chen},\ and\ \citenamefont {Wu}}]{LCO2018}%
  \BibitemOpen
  \bibfield  {author} {\bibinfo {author} {\bibfnamefont {S.}~\bibnamefont {Zhu}}, \bibinfo {author} {\bibfnamefont {D.}~\bibnamefont {Meng}}, \bibinfo {author} {\bibfnamefont {G.}~\bibnamefont {Liang}}, \bibinfo {author} {\bibfnamefont {G.}~\bibnamefont {Shi}}, \bibinfo {author} {\bibfnamefont {P.}~\bibnamefont {Zhao}}, \bibinfo {author} {\bibfnamefont {P.}~\bibnamefont {Cheng}}, \bibinfo {author} {\bibfnamefont {Y.}~\bibnamefont {Li}}, \bibinfo {author} {\bibfnamefont {X.}~\bibnamefont {Zhai}}, \bibinfo {author} {\bibfnamefont {Y.}~\bibnamefont {Lu}}, \bibinfo {author} {\bibfnamefont {L.}~\bibnamefont {Chen}},\ and\ \bibinfo {author} {\bibfnamefont {K.}~\bibnamefont {Wu}},\ }\bibfield  {title} {\bibinfo {title} {Proximity-induced magnetism and an anomalous hall effect in bi2se3/lacoo3: a topological insulator/ferromagnetic insulator thin film heterostructure},\ }\href {https://doi.org/10.1039/C8NR02083C} {\bibfield  {journal} {\bibinfo  {journal} {Nanoscale}\ }\textbf {\bibinfo {volume} {10}},\ \bibinfo
  {pages} {10041} (\bibinfo {year} {2018})}\BibitemShut {NoStop}%
\bibitem [{\citenamefont {Wang}\ \emph {et~al.}(2020)\citenamefont {Wang}, \citenamefont {Liu}, \citenamefont {Wu}, \citenamefont {Hou}, \citenamefont {Jiang}, \citenamefont {Li}, \citenamefont {Pandey}, \citenamefont {Chen}, \citenamefont {Yang}, \citenamefont {Wang}, \citenamefont {Wei}, \citenamefont {Lei}, \citenamefont {Kang}, \citenamefont {Wen}, \citenamefont {Nie}, \citenamefont {Zhao},\ and\ \citenamefont {Wang}}]{FGT2020}%
  \BibitemOpen
  \bibfield  {author} {\bibinfo {author} {\bibfnamefont {H.}~\bibnamefont {Wang}}, \bibinfo {author} {\bibfnamefont {Y.}~\bibnamefont {Liu}}, \bibinfo {author} {\bibfnamefont {P.}~\bibnamefont {Wu}}, \bibinfo {author} {\bibfnamefont {W.}~\bibnamefont {Hou}}, \bibinfo {author} {\bibfnamefont {Y.}~\bibnamefont {Jiang}}, \bibinfo {author} {\bibfnamefont {X.}~\bibnamefont {Li}}, \bibinfo {author} {\bibfnamefont {C.}~\bibnamefont {Pandey}}, \bibinfo {author} {\bibfnamefont {D.}~\bibnamefont {Chen}}, \bibinfo {author} {\bibfnamefont {Q.}~\bibnamefont {Yang}}, \bibinfo {author} {\bibfnamefont {H.}~\bibnamefont {Wang}}, \bibinfo {author} {\bibfnamefont {D.}~\bibnamefont {Wei}}, \bibinfo {author} {\bibfnamefont {N.}~\bibnamefont {Lei}}, \bibinfo {author} {\bibfnamefont {W.}~\bibnamefont {Kang}}, \bibinfo {author} {\bibfnamefont {L.}~\bibnamefont {Wen}}, \bibinfo {author} {\bibfnamefont {T.}~\bibnamefont {Nie}}, \bibinfo {author} {\bibfnamefont {W.}~\bibnamefont {Zhao}},\ and\ \bibinfo {author} {\bibfnamefont {K.~L.}\
  \bibnamefont {Wang}},\ }\bibfield  {title} {\bibinfo {title} {{Above Room-Temperature Ferromagnetism in Wafer-Scale Two-Dimensional van der Waals Fe$_3$GeTe$_2$ Tailored by a Topological Insulator}},\ }\href {https://doi.org/10.1021/acsnano.0c03152} {\bibfield  {journal} {\bibinfo  {journal} {ACS Nano}\ }\textbf {\bibinfo {volume} {14}},\ \bibinfo {pages} {10045} (\bibinfo {year} {2020})}\BibitemShut {NoStop}%
\bibitem [{\citenamefont {Miao}\ \emph {et~al.}(2023)\citenamefont {Miao}, \citenamefont {Kang}, \citenamefont {Song},\ and\ \citenamefont {Zhang}}]{SRO2023}%
  \BibitemOpen
  \bibfield  {author} {\bibinfo {author} {\bibfnamefont {Q.}~\bibnamefont {Miao}}, \bibinfo {author} {\bibfnamefont {C.}~\bibnamefont {Kang}}, \bibinfo {author} {\bibfnamefont {Y.-H.}\ \bibnamefont {Song}},\ and\ \bibinfo {author} {\bibfnamefont {W.}~\bibnamefont {Zhang}},\ }\bibfield  {title} {\bibinfo {title} {{Magnetic proximity effect in the heterostructures of topological insulators and SrRuO3}},\ }\href {https://doi.org/10.1063/5.0147158} {\bibfield  {journal} {\bibinfo  {journal} {Appl. Phys. Lett.}\ }\textbf {\bibinfo {volume} {122}},\ \bibinfo {pages} {183103} (\bibinfo {year} {2023})}\BibitemShut {NoStop}%
\bibitem [{\citenamefont {Kim}\ \emph {et~al.}(2017)\citenamefont {Kim}, \citenamefont {Kim}, \citenamefont {Wang}, \citenamefont {Sinova},\ and\ \citenamefont {Wu}}]{Kim2017DFT}%
  \BibitemOpen
  \bibfield  {author} {\bibinfo {author} {\bibfnamefont {J.}~\bibnamefont {Kim}}, \bibinfo {author} {\bibfnamefont {K.-W.}\ \bibnamefont {Kim}}, \bibinfo {author} {\bibfnamefont {H.}~\bibnamefont {Wang}}, \bibinfo {author} {\bibfnamefont {J.}~\bibnamefont {Sinova}},\ and\ \bibinfo {author} {\bibfnamefont {R.}~\bibnamefont {Wu}},\ }\bibfield  {title} {\bibinfo {title} {Understanding the giant enhancement of exchange interaction in ${\mathrm{bi}}_{2}{\mathrm{se}}_{3}\text{\ensuremath{-}}\mathrm{EuS}$ heterostructures},\ }\href {https://doi.org/10.1103/PhysRevLett.119.027201} {\bibfield  {journal} {\bibinfo  {journal} {Phys. Rev. Lett.}\ }\textbf {\bibinfo {volume} {119}},\ \bibinfo {pages} {027201} (\bibinfo {year} {2017})}\BibitemShut {NoStop}%
\bibitem [{\citenamefont {Zhong}\ \emph {et~al.}(2023)\citenamefont {Zhong}, \citenamefont {Peng}, \citenamefont {Huang},\ and\ \citenamefont {et~al.}}]{shen2023}%
  \BibitemOpen
  \bibfield  {author} {\bibinfo {author} {\bibfnamefont {Y.}~\bibnamefont {Zhong}}, \bibinfo {author} {\bibfnamefont {C.}~\bibnamefont {Peng}}, \bibinfo {author} {\bibfnamefont {H.}~\bibnamefont {Huang}},\ and\ \bibinfo {author} {\bibnamefont {et~al.}},\ }\bibfield  {title} {\bibinfo {title} {From stoner to local moment magnetism in atomically thin {Cr$_2$Te$_3$}},\ }\href {https://doi.org/10.1038/s41467-023-40997-1} {\bibfield  {journal} {\bibinfo  {journal} {Nature Communications}\ }\textbf {\bibinfo {volume} {14}},\ \bibinfo {pages} {5340} (\bibinfo {year} {2023})}\BibitemShut {NoStop}%
\bibitem [{\citenamefont {Ruderman}\ and\ \citenamefont {Kittel}(1954)}]{RK1954}%
  \BibitemOpen
  \bibfield  {author} {\bibinfo {author} {\bibfnamefont {M.~A.}\ \bibnamefont {Ruderman}}\ and\ \bibinfo {author} {\bibfnamefont {C.}~\bibnamefont {Kittel}},\ }\bibfield  {title} {\bibinfo {title} {Indirect exchange coupling of nuclear magnetic moments by conduction electrons},\ }\href {https://doi.org/10.1103/PhysRev.96.99} {\bibfield  {journal} {\bibinfo  {journal} {Phys. Rev.}\ }\textbf {\bibinfo {volume} {96}},\ \bibinfo {pages} {99} (\bibinfo {year} {1954})}\BibitemShut {NoStop}%
\bibitem [{\citenamefont {Kasuya}(1956)}]{Kasuya56}%
  \BibitemOpen
  \bibfield  {author} {\bibinfo {author} {\bibfnamefont {T.}~\bibnamefont {Kasuya}},\ }\bibfield  {title} {\bibinfo {title} {{A Theory of Metallic Ferro- and Antiferromagnetism on Zener's Model}},\ }\href {https://doi.org/10.1143/PTP.16.45} {\bibfield  {journal} {\bibinfo  {journal} {Progress of Theoretical Physics}\ }\textbf {\bibinfo {volume} {16}},\ \bibinfo {pages} {45} (\bibinfo {year} {1956})}\BibitemShut {NoStop}%
\bibitem [{\citenamefont {Yosida}(1957)}]{Yoshida57}%
  \BibitemOpen
  \bibfield  {author} {\bibinfo {author} {\bibfnamefont {K.}~\bibnamefont {Yosida}},\ }\bibfield  {title} {\bibinfo {title} {Magnetic properties of cu-mn alloys},\ }\href {https://doi.org/10.1103/PhysRev.106.893} {\bibfield  {journal} {\bibinfo  {journal} {Phys. Rev.}\ }\textbf {\bibinfo {volume} {106}},\ \bibinfo {pages} {893} (\bibinfo {year} {1957})}\BibitemShut {NoStop}%
\bibitem [{\citenamefont {Bloembergen}\ and\ \citenamefont {Rowland}(1955)}]{BR1955}%
  \BibitemOpen
  \bibfield  {author} {\bibinfo {author} {\bibfnamefont {N.}~\bibnamefont {Bloembergen}}\ and\ \bibinfo {author} {\bibfnamefont {T.~J.}\ \bibnamefont {Rowland}},\ }\bibfield  {title} {\bibinfo {title} {{Nuclear Spin Exchange in Solids: Tl$^{203}$ and Tl$^{205}$ Magnetic Resonance in Thallium and Thallic Oxide}},\ }\href {https://doi.org/10.1103/PhysRev.97.1679} {\bibfield  {journal} {\bibinfo  {journal} {Physical Review}\ }\textbf {\bibinfo {volume} {97}},\ \bibinfo {pages} {1679} (\bibinfo {year} {1955})}\BibitemShut {NoStop}%
\bibitem [{\citenamefont {Liu}\ \emph {et~al.}(2009)\citenamefont {Liu}, \citenamefont {Liu}, \citenamefont {Xu}, \citenamefont {Qi},\ and\ \citenamefont {Zhang}}]{Zhang2009}%
  \BibitemOpen
  \bibfield  {author} {\bibinfo {author} {\bibfnamefont {Q.}~\bibnamefont {Liu}}, \bibinfo {author} {\bibfnamefont {C.-X.}\ \bibnamefont {Liu}}, \bibinfo {author} {\bibfnamefont {C.}~\bibnamefont {Xu}}, \bibinfo {author} {\bibfnamefont {X.-L.}\ \bibnamefont {Qi}},\ and\ \bibinfo {author} {\bibfnamefont {S.-C.}\ \bibnamefont {Zhang}},\ }\bibfield  {title} {\bibinfo {title} {Magnetic impurities on the surface of a topological insulator},\ }\href {https://doi.org/10.1103/PhysRevLett.102.156603} {\bibfield  {journal} {\bibinfo  {journal} {Phys. Rev. Lett.}\ }\textbf {\bibinfo {volume} {102}},\ \bibinfo {pages} {156603} (\bibinfo {year} {2009})}\BibitemShut {NoStop}%
\bibitem [{\citenamefont {Ye}\ \emph {et~al.}(2010)\citenamefont {Ye}, \citenamefont {Ding}, \citenamefont {Zhai},\ and\ \citenamefont {Su}}]{Ye_2010}%
  \BibitemOpen
  \bibfield  {author} {\bibinfo {author} {\bibfnamefont {F.}~\bibnamefont {Ye}}, \bibinfo {author} {\bibfnamefont {G.~H.}\ \bibnamefont {Ding}}, \bibinfo {author} {\bibfnamefont {H.}~\bibnamefont {Zhai}},\ and\ \bibinfo {author} {\bibfnamefont {Z.~B.}\ \bibnamefont {Su}},\ }\bibfield  {title} {\bibinfo {title} {Spin helix of magnetic impurities in two-dimensional helical metal},\ }\href {https://doi.org/10.1209/0295-5075/90/47001} {\bibfield  {journal} {\bibinfo  {journal} {Europhysics Letters}\ }\textbf {\bibinfo {volume} {90}},\ \bibinfo {pages} {47001} (\bibinfo {year} {2010})}\BibitemShut {NoStop}%
\bibitem [{\citenamefont {Biswas}\ and\ \citenamefont {Balatsky}(2010)}]{Biswas2010BR}%
  \BibitemOpen
  \bibfield  {author} {\bibinfo {author} {\bibfnamefont {R.~R.}\ \bibnamefont {Biswas}}\ and\ \bibinfo {author} {\bibfnamefont {A.~V.}\ \bibnamefont {Balatsky}},\ }\bibfield  {title} {\bibinfo {title} {Impurity-induced states on the surface of three-dimensional topological insulators},\ }\href {https://doi.org/10.1103/PhysRevB.81.233405} {\bibfield  {journal} {\bibinfo  {journal} {Phys. Rev. B}\ }\textbf {\bibinfo {volume} {81}},\ \bibinfo {pages} {233405} (\bibinfo {year} {2010})}\BibitemShut {NoStop}%
\bibitem [{\citenamefont {Garate}\ and\ \citenamefont {Franz}(2010)}]{Garate2010RKKY}%
  \BibitemOpen
  \bibfield  {author} {\bibinfo {author} {\bibfnamefont {I.}~\bibnamefont {Garate}}\ and\ \bibinfo {author} {\bibfnamefont {M.}~\bibnamefont {Franz}},\ }\bibfield  {title} {\bibinfo {title} {Magnetoelectric response of the time-reversal invariant helical metal},\ }\href {https://doi.org/10.1103/PhysRevB.81.172408} {\bibfield  {journal} {\bibinfo  {journal} {Phys. Rev. B}\ }\textbf {\bibinfo {volume} {81}},\ \bibinfo {pages} {172408} (\bibinfo {year} {2010})}\BibitemShut {NoStop}%
\bibitem [{\citenamefont {Zhu}\ \emph {et~al.}(2011)\citenamefont {Zhu}, \citenamefont {Yao}, \citenamefont {Zhang},\ and\ \citenamefont {Chang}}]{Chang2011}%
  \BibitemOpen
  \bibfield  {author} {\bibinfo {author} {\bibfnamefont {J.-J.}\ \bibnamefont {Zhu}}, \bibinfo {author} {\bibfnamefont {D.-X.}\ \bibnamefont {Yao}}, \bibinfo {author} {\bibfnamefont {S.-C.}\ \bibnamefont {Zhang}},\ and\ \bibinfo {author} {\bibfnamefont {K.}~\bibnamefont {Chang}},\ }\bibfield  {title} {\bibinfo {title} {{Electrically Controllable Surface Magnetism on the Surface of Topological Insulators}},\ }\href {https://doi.org/10.1103/PhysRevLett.106.097201} {\bibfield  {journal} {\bibinfo  {journal} {Physical Review Letters}\ }\textbf {\bibinfo {volume} {106}},\ \bibinfo {pages} {097201} (\bibinfo {year} {2011})}\BibitemShut {NoStop}%
\bibitem [{\citenamefont {Efimkin}\ and\ \citenamefont {Galitski}(2014)}]{Galitski2014}%
  \BibitemOpen
  \bibfield  {author} {\bibinfo {author} {\bibfnamefont {D.~K.}\ \bibnamefont {Efimkin}}\ and\ \bibinfo {author} {\bibfnamefont {V.}~\bibnamefont {Galitski}},\ }\bibfield  {title} {\bibinfo {title} {{Self-consistent theory of ferromagnetism on the surface of a topological insulator}},\ }\href {https://doi.org/10.1103/PhysRevB.89.115431} {\bibfield  {journal} {\bibinfo  {journal} {Physical Review B}\ }\textbf {\bibinfo {volume} {89}},\ \bibinfo {pages} {115431} (\bibinfo {year} {2014})}\BibitemShut {NoStop}%
\bibitem [{\citenamefont {Zyuzin}\ and\ \citenamefont {Loss}(2014)}]{Zyuzin2014}%
  \BibitemOpen
  \bibfield  {author} {\bibinfo {author} {\bibfnamefont {A.~A.}\ \bibnamefont {Zyuzin}}\ and\ \bibinfo {author} {\bibfnamefont {D.}~\bibnamefont {Loss}},\ }\bibfield  {title} {\bibinfo {title} {Rkky interaction on surfaces of topological insulators with superconducting proximity effect},\ }\href {https://doi.org/10.1103/PhysRevB.90.125443} {\bibfield  {journal} {\bibinfo  {journal} {Phys. Rev. B}\ }\textbf {\bibinfo {volume} {90}},\ \bibinfo {pages} {125443} (\bibinfo {year} {2014})}\BibitemShut {NoStop}%
\bibitem [{\citenamefont {Hasan}\ and\ \citenamefont {Kane}(2010)}]{Hasan_20103DTI}%
  \BibitemOpen
  \bibfield  {author} {\bibinfo {author} {\bibfnamefont {M.~Z.}\ \bibnamefont {Hasan}}\ and\ \bibinfo {author} {\bibfnamefont {C.~L.}\ \bibnamefont {Kane}},\ }\bibfield  {title} {\bibinfo {title} {Colloquium: Topological insulators},\ }\href {https://doi.org/10.1103/revmodphys.82.3045} {\bibfield  {journal} {\bibinfo  {journal} {Reviews of Modern Physics}\ }\textbf {\bibinfo {volume} {82}},\ \bibinfo {pages} {3045–3067} (\bibinfo {year} {2010})}\BibitemShut {NoStop}%
\bibitem [{\citenamefont {Qi}\ and\ \citenamefont {Zhang}(2011)}]{Qi3DTI}%
  \BibitemOpen
  \bibfield  {author} {\bibinfo {author} {\bibfnamefont {X.-L.}\ \bibnamefont {Qi}}\ and\ \bibinfo {author} {\bibfnamefont {S.-C.}\ \bibnamefont {Zhang}},\ }\bibfield  {title} {\bibinfo {title} {Topological insulators and superconductors},\ }\href {https://doi.org/10.1103/RevModPhys.83.1057} {\bibfield  {journal} {\bibinfo  {journal} {Rev. Mod. Phys.}\ }\textbf {\bibinfo {volume} {83}},\ \bibinfo {pages} {1057} (\bibinfo {year} {2011})}\BibitemShut {NoStop}%
\bibitem [{\citenamefont {Abanin}\ and\ \citenamefont {Pesin}(2011)}]{Pesin2011}%
  \BibitemOpen
  \bibfield  {author} {\bibinfo {author} {\bibfnamefont {D.~A.}\ \bibnamefont {Abanin}}\ and\ \bibinfo {author} {\bibfnamefont {D.~A.}\ \bibnamefont {Pesin}},\ }\bibfield  {title} {\bibinfo {title} {{Ordering of Magnetic Impurities and Tunable Electronic Properties of Topological Insulators}},\ }\href {https://doi.org/10.1103/PhysRevLett.106.136802} {\bibfield  {journal} {\bibinfo  {journal} {Physical Review Letters}\ }\textbf {\bibinfo {volume} {106}},\ \bibinfo {pages} {136802} (\bibinfo {year} {2011})}\BibitemShut {NoStop}%
\bibitem [{\citenamefont {Luo}\ and\ \citenamefont {Qi}(2013)}]{Luo2013Gap}%
  \BibitemOpen
  \bibfield  {author} {\bibinfo {author} {\bibfnamefont {W.}~\bibnamefont {Luo}}\ and\ \bibinfo {author} {\bibfnamefont {X.-L.}\ \bibnamefont {Qi}},\ }\bibfield  {title} {\bibinfo {title} {Massive dirac surface states in topological insulator/magnetic insulator heterostructures},\ }\href {https://doi.org/10.1103/PhysRevB.87.085431} {\bibfield  {journal} {\bibinfo  {journal} {Phys. Rev. B}\ }\textbf {\bibinfo {volume} {87}},\ \bibinfo {pages} {085431} (\bibinfo {year} {2013})}\BibitemShut {NoStop}%
\bibitem [{\citenamefont {He}\ \emph {et~al.}(2015)\citenamefont {He}, \citenamefont {Li}, \citenamefont {Chen},\ and\ \citenamefont {Wu}}]{BST2015}%
  \BibitemOpen
  \bibfield  {author} {\bibinfo {author} {\bibfnamefont {X.}~\bibnamefont {He}}, \bibinfo {author} {\bibfnamefont {H.}~\bibnamefont {Li}}, \bibinfo {author} {\bibfnamefont {L.}~\bibnamefont {Chen}},\ and\ \bibinfo {author} {\bibfnamefont {K.}~\bibnamefont {Wu}},\ }\bibfield  {title} {\bibinfo {title} {{Substitution-induced spin-splitted surface states in topological insulator (Bi$_{1-x}$Sb$_x$)$_2$Te$_3$}},\ }\href {https://doi.org/10.1038/srep08830} {\bibfield  {journal} {\bibinfo  {journal} {Scientific Reports}\ }\textbf {\bibinfo {volume} {5}},\ \bibinfo {pages} {8830} (\bibinfo {year} {2015})}\BibitemShut {NoStop}%
\bibitem [{\citenamefont {Vleck}(1932)}]{VanVleck1932}%
  \BibitemOpen
  \bibfield  {author} {\bibinfo {author} {\bibfnamefont {J.~H.~V.}\ \bibnamefont {Vleck}},\ }\href@noop {} {\emph {\bibinfo {title} {The Theory of Electric and Magnetic Susceptibilities}}}\ (\bibinfo  {publisher} {Oxford University Press},\ \bibinfo {address} {London},\ \bibinfo {year} {1932})\BibitemShut {NoStop}%
\bibitem [{\citenamefont {Wang}\ \emph {et~al.}(2015)\citenamefont {Wang}, \citenamefont {Lian},\ and\ \citenamefont {Zhang}}]{Wang2015}%
  \BibitemOpen
  \bibfield  {author} {\bibinfo {author} {\bibfnamefont {J.}~\bibnamefont {Wang}}, \bibinfo {author} {\bibfnamefont {B.}~\bibnamefont {Lian}},\ and\ \bibinfo {author} {\bibfnamefont {S.-C.}\ \bibnamefont {Zhang}},\ }\bibfield  {title} {\bibinfo {title} {Electrically tunable magnetism in magnetic topological insulators},\ }\href {https://doi.org/10.1103/PhysRevLett.115.036805} {\bibfield  {journal} {\bibinfo  {journal} {Phys. Rev. Lett.}\ }\textbf {\bibinfo {volume} {115}},\ \bibinfo {pages} {036805} (\bibinfo {year} {2015})}\BibitemShut {NoStop}%
\bibitem [{\citenamefont {Zhang}\ \emph {et~al.}(2010)\citenamefont {Zhang}, \citenamefont {He}, \citenamefont {Chang}, \citenamefont {Song}, \citenamefont {Wang}, \citenamefont {Chen}, \citenamefont {Jia}, \citenamefont {Fang}, \citenamefont {Dai}, \citenamefont {Shan}, \citenamefont {Shen}, \citenamefont {Niu}, \citenamefont {Qi}, \citenamefont {Zhang}, \citenamefont {Ma},\ and\ \citenamefont {Xue}}]{Zhang20106QL}%
  \BibitemOpen
  \bibfield  {author} {\bibinfo {author} {\bibfnamefont {Y.}~\bibnamefont {Zhang}}, \bibinfo {author} {\bibfnamefont {K.}~\bibnamefont {He}}, \bibinfo {author} {\bibfnamefont {C.-Z.}\ \bibnamefont {Chang}}, \bibinfo {author} {\bibfnamefont {C.-L.}\ \bibnamefont {Song}}, \bibinfo {author} {\bibfnamefont {L.-L.}\ \bibnamefont {Wang}}, \bibinfo {author} {\bibfnamefont {X.}~\bibnamefont {Chen}}, \bibinfo {author} {\bibfnamefont {J.-F.}\ \bibnamefont {Jia}}, \bibinfo {author} {\bibfnamefont {Z.}~\bibnamefont {Fang}}, \bibinfo {author} {\bibfnamefont {X.}~\bibnamefont {Dai}}, \bibinfo {author} {\bibfnamefont {W.-Y.}\ \bibnamefont {Shan}}, \bibinfo {author} {\bibfnamefont {S.-Q.}\ \bibnamefont {Shen}}, \bibinfo {author} {\bibfnamefont {Q.}~\bibnamefont {Niu}}, \bibinfo {author} {\bibfnamefont {X.-L.}\ \bibnamefont {Qi}}, \bibinfo {author} {\bibfnamefont {S.-C.}\ \bibnamefont {Zhang}}, \bibinfo {author} {\bibfnamefont {X.-C.}\ \bibnamefont {Ma}},\ and\ \bibinfo {author} {\bibfnamefont {Q.-K.}\ \bibnamefont {Xue}},\
  }\bibfield  {title} {\bibinfo {title} {Crossover of the three-dimensional topological insulator bi2se3 to the two-dimensional limit},\ }\href {https://doi.org/10.1038/nphys1689} {\bibfield  {journal} {\bibinfo  {journal} {Nature Physics}\ }\textbf {\bibinfo {volume} {6}},\ \bibinfo {pages} {584–588} (\bibinfo {year} {2010})}\BibitemShut {NoStop}%
\bibitem [{\citenamefont {Shan}\ \emph {et~al.}(2010)\citenamefont {Shan}, \citenamefont {Lu},\ and\ \citenamefont {Shen}}]{Shan_2010}%
  \BibitemOpen
  \bibfield  {author} {\bibinfo {author} {\bibfnamefont {W.-Y.}\ \bibnamefont {Shan}}, \bibinfo {author} {\bibfnamefont {H.-Z.}\ \bibnamefont {Lu}},\ and\ \bibinfo {author} {\bibfnamefont {S.-Q.}\ \bibnamefont {Shen}},\ }\bibfield  {title} {\bibinfo {title} {Effective continuous model for surface states and thin films of three-dimensional topological insulators},\ }\href {https://doi.org/10.1088/1367-2630/12/4/043048} {\bibfield  {journal} {\bibinfo  {journal} {New Journal of Physics}\ }\textbf {\bibinfo {volume} {12}},\ \bibinfo {pages} {043048} (\bibinfo {year} {2010})}\BibitemShut {NoStop}%
\bibitem [{\citenamefont {Bernevig}\ \emph {et~al.}(2006)\citenamefont {Bernevig}, \citenamefont {Hughes},\ and\ \citenamefont {Zhang}}]{BHZ2006}%
  \BibitemOpen
  \bibfield  {author} {\bibinfo {author} {\bibfnamefont {B.~A.}\ \bibnamefont {Bernevig}}, \bibinfo {author} {\bibfnamefont {T.~L.}\ \bibnamefont {Hughes}},\ and\ \bibinfo {author} {\bibfnamefont {S.-C.}\ \bibnamefont {Zhang}},\ }\bibfield  {title} {\bibinfo {title} {Quantum spin hall effect and topological phase transition in hgte quantum wells},\ }\href {https://doi.org/10.1126/science.1133734} {\bibfield  {journal} {\bibinfo  {journal} {Science}\ }\textbf {\bibinfo {volume} {314}},\ \bibinfo {pages} {1757} (\bibinfo {year} {2006})}\BibitemShut {NoStop}%
\bibitem [{\citenamefont {Liu}\ \emph {et~al.}(2010)\citenamefont {Liu}, \citenamefont {Zhang}, \citenamefont {Yan}, \citenamefont {Qi}, \citenamefont {Frauenheim}, \citenamefont {Dai}, \citenamefont {Fang},\ and\ \citenamefont {Zhang}}]{Liu2010Oscillatory}%
  \BibitemOpen
  \bibfield  {author} {\bibinfo {author} {\bibfnamefont {C.-X.}\ \bibnamefont {Liu}}, \bibinfo {author} {\bibfnamefont {H.}~\bibnamefont {Zhang}}, \bibinfo {author} {\bibfnamefont {B.}~\bibnamefont {Yan}}, \bibinfo {author} {\bibfnamefont {X.-L.}\ \bibnamefont {Qi}}, \bibinfo {author} {\bibfnamefont {T.}~\bibnamefont {Frauenheim}}, \bibinfo {author} {\bibfnamefont {X.}~\bibnamefont {Dai}}, \bibinfo {author} {\bibfnamefont {Z.}~\bibnamefont {Fang}},\ and\ \bibinfo {author} {\bibfnamefont {S.-C.}\ \bibnamefont {Zhang}},\ }\bibfield  {title} {\bibinfo {title} {Oscillatory crossover from two-dimensional to three-dimensional topological insulators},\ }\href {https://doi.org/10.1103/PhysRevB.81.041307} {\bibfield  {journal} {\bibinfo  {journal} {Phys. Rev. B}\ }\textbf {\bibinfo {volume} {81}},\ \bibinfo {pages} {041307} (\bibinfo {year} {2010})}\BibitemShut {NoStop}%
\bibitem [{\citenamefont {Abrikosov}\ \emph {et~al.}(1965)\citenamefont {Abrikosov}, \citenamefont {Gorkov},\ and\ \citenamefont {Dzyaloshinskii}}]{AGD1965Pergamon}%
  \BibitemOpen
  \bibfield  {author} {\bibinfo {author} {\bibfnamefont {A.~A.}\ \bibnamefont {Abrikosov}}, \bibinfo {author} {\bibfnamefont {L.~P.}\ \bibnamefont {Gorkov}},\ and\ \bibinfo {author} {\bibfnamefont {I.~E.}\ \bibnamefont {Dzyaloshinskii}},\ }\href@noop {} {\emph {\bibinfo {title} {Methods of Quantum Field Theory in Statistical Physics}}}\ (\bibinfo  {publisher} {Pergamon Press},\ \bibinfo {address} {Oxford},\ \bibinfo {year} {1965})\BibitemShut {NoStop}%
\bibitem [{\citenamefont {Coleman}(2015)}]{Coleman2015}%
  \BibitemOpen
  \bibfield  {author} {\bibinfo {author} {\bibfnamefont {P.}~\bibnamefont {Coleman}},\ }\href@noop {} {\emph {\bibinfo {title} {Introduction to Many-Body Physics}}}\ (\bibinfo  {publisher} {Cambridge University Press},\ \bibinfo {address} {Cambridge, UK},\ \bibinfo {year} {2015})\BibitemShut {NoStop}%
\bibitem [{\citenamefont {Saremi}(2007)}]{Saremi2007}%
  \BibitemOpen
  \bibfield  {author} {\bibinfo {author} {\bibfnamefont {S.}~\bibnamefont {Saremi}},\ }\bibfield  {title} {\bibinfo {title} {Rkky in half-filled bipartite lattices: Graphene as an example},\ }\href {https://doi.org/10.1103/PhysRevB.76.184430} {\bibfield  {journal} {\bibinfo  {journal} {Phys. Rev. B}\ }\textbf {\bibinfo {volume} {76}},\ \bibinfo {pages} {184430} (\bibinfo {year} {2007})}\BibitemShut {NoStop}%
\bibitem [{\citenamefont {Liu}\ \emph {et~al.}(2016)\citenamefont {Liu}, \citenamefont {Roy},\ and\ \citenamefont {Sau}}]{Sau2016}%
  \BibitemOpen
  \bibfield  {author} {\bibinfo {author} {\bibfnamefont {C.-X.}\ \bibnamefont {Liu}}, \bibinfo {author} {\bibfnamefont {B.}~\bibnamefont {Roy}},\ and\ \bibinfo {author} {\bibfnamefont {J.~D.}\ \bibnamefont {Sau}},\ }\bibfield  {title} {\bibinfo {title} {Ferromagnetism and glassiness on the surface of topological insulators},\ }\href {https://doi.org/10.1103/PhysRevB.94.235421} {\bibfield  {journal} {\bibinfo  {journal} {Phys. Rev. B}\ }\textbf {\bibinfo {volume} {94}},\ \bibinfo {pages} {235421} (\bibinfo {year} {2016})}\BibitemShut {NoStop}%
\bibitem [{\citenamefont {Dugaev}\ \emph {et~al.}(1994)\citenamefont {Dugaev}, \citenamefont {Litvinov},\ and\ \citenamefont {Petrov}}]{dugaev1994GFs}%
  \BibitemOpen
  \bibfield  {author} {\bibinfo {author} {\bibfnamefont {V.}~\bibnamefont {Dugaev}}, \bibinfo {author} {\bibfnamefont {V.}~\bibnamefont {Litvinov}},\ and\ \bibinfo {author} {\bibfnamefont {P.}~\bibnamefont {Petrov}},\ }\bibfield  {title} {\bibinfo {title} {Magnetic impurity interactions in a quantum well on the base of iv-vi semiconductors},\ }\href {https://doi.org/https://doi.org/10.1006/spmi.1994.1160} {\bibfield  {journal} {\bibinfo  {journal} {Superlattices and Microstructures}\ }\textbf {\bibinfo {volume} {16}},\ \bibinfo {pages} {413} (\bibinfo {year} {1994})}\BibitemShut {NoStop}%
\bibitem [{\citenamefont {Gradshteyn}\ and\ \citenamefont {Ryzhik}(2014)}]{gradshteyn2014table}%
  \BibitemOpen
  \bibfield  {author} {\bibinfo {author} {\bibfnamefont {I.}~\bibnamefont {Gradshteyn}}\ and\ \bibinfo {author} {\bibfnamefont {I.}~\bibnamefont {Ryzhik}},\ }\href@noop {} {\emph {\bibinfo {title} {Table of Integrals, Series, and Products}}}\ (\bibinfo  {publisher} {Elsevier Science},\ \bibinfo {year} {2014})\BibitemShut {NoStop}%
\bibitem [{\citenamefont {Yarmohammadi}\ \emph {et~al.}(2023)\citenamefont {Yarmohammadi}, \citenamefont {Bukov},\ and\ \citenamefont {Kolodrubetz}}]{Yarmohammadi2023}%
  \BibitemOpen
  \bibfield  {author} {\bibinfo {author} {\bibfnamefont {M.}~\bibnamefont {Yarmohammadi}}, \bibinfo {author} {\bibfnamefont {M.}~\bibnamefont {Bukov}},\ and\ \bibinfo {author} {\bibfnamefont {M.~H.}\ \bibnamefont {Kolodrubetz}},\ }\bibfield  {title} {\bibinfo {title} {Noncollinear twisted rkky interaction on the optically driven snte(001) surface},\ }\href {https://doi.org/10.1103/PhysRevB.107.054439} {\bibfield  {journal} {\bibinfo  {journal} {Phys. Rev. B}\ }\textbf {\bibinfo {volume} {107}},\ \bibinfo {pages} {054439} (\bibinfo {year} {2023})}\BibitemShut {NoStop}%
\bibitem [{\citenamefont {Landau}\ and\ \citenamefont {Lifshitz}(1935)}]{landau1935theory}%
  \BibitemOpen
  \bibfield  {author} {\bibinfo {author} {\bibfnamefont {L.}~\bibnamefont {Landau}}\ and\ \bibinfo {author} {\bibfnamefont {E.}~\bibnamefont {Lifshitz}},\ }\bibfield  {title} {\bibinfo {title} {On the theory of the dispersion of magnetic permeability in ferromagnetic bodies. reproduced in collected papers of ld landau},\ }\href@noop {} {\bibfield  {journal} {\bibinfo  {journal} {Pergamon, New York}\ } (\bibinfo {year} {1935})}\BibitemShut {NoStop}%
\bibitem [{\citenamefont {Shiranzaei}\ \emph {et~al.}(2017)\citenamefont {Shiranzaei}, \citenamefont {Cheraghchi},\ and\ \citenamefont {Parhizgar}}]{Shiranzaei2017}%
  \BibitemOpen
  \bibfield  {author} {\bibinfo {author} {\bibfnamefont {M.}~\bibnamefont {Shiranzaei}}, \bibinfo {author} {\bibfnamefont {H.}~\bibnamefont {Cheraghchi}},\ and\ \bibinfo {author} {\bibfnamefont {F.}~\bibnamefont {Parhizgar}},\ }\bibfield  {title} {\bibinfo {title} {Effect of the rashba splitting on the rkky interaction in topological-insulator thin films},\ }\href {https://doi.org/10.1103/PhysRevB.96.024413} {\bibfield  {journal} {\bibinfo  {journal} {Phys. Rev. B}\ }\textbf {\bibinfo {volume} {96}},\ \bibinfo {pages} {024413} (\bibinfo {year} {2017})}\BibitemShut {NoStop}%
\bibitem [{\citenamefont {Litvinov}(2016)}]{LitvinovBROscillatory}%
  \BibitemOpen
  \bibfield  {author} {\bibinfo {author} {\bibfnamefont {V.~I.}\ \bibnamefont {Litvinov}},\ }\bibfield  {title} {\bibinfo {title} {Oscillating bloembergen-rowland interaction in three-dimensional topological insulators},\ }\href {https://doi.org/10.1103/PhysRevB.94.035138} {\bibfield  {journal} {\bibinfo  {journal} {Phys. Rev. B}\ }\textbf {\bibinfo {volume} {94}},\ \bibinfo {pages} {035138} (\bibinfo {year} {2016})}\BibitemShut {NoStop}%
\end{thebibliography}%

\end{document}